\documentclass[11pt]{article}

\newsavebox{\foobox}
\newcommand{\slantbox}[2][0]{\mbox{%
        \sbox{\foobox}{#2}%
        \hskip\wd\foobox
        \pdfsave
        \pdfsetmatrix{1 0 #1 1}%
        \llap{\usebox{\foobox}}%
        \pdfrestore
}}
\newcommand\unslant[2][-.25]{\slantbox[#1]{$#2$}}

\newcommand{\mpi}{\text{\unslant[-.18]\pi}}
\newcommand{\mdelta}{\text{\unslant[-.18]\delta}}

\usepackage[left=2cm, right=2cm, top=2.5cm, bottom=2.5cm]{geometry}
\usepackage[x11names]{xcolor}
\usepackage{fancyhdr, amssymb, cancel, amsmath, graphicx, pgfplots, tikz}
\usepackage{isomath}

\usetikzlibrary{shadows}

\newcommand{\stylecolor}{blue!50!black}

\usepackage[labelfont={bf,sf, color=\stylecolor}, margin={1.5cm,0cm}]{caption}

\usepackage[colorlinks=true, urlcolor=\stylecolor!70!white, linkcolor=\stylecolor, citecolor=\stylecolor!70!white, hyperindex=true, linktocpage=true]{hyperref}

\usepackage[explicit]{titlesec}

\newcommand*\sectionlabel{}
\titleformat{\section}
  {\gdef\sectionlabel{}
   \Large\bfseries\scshape}
  {\gdef\sectionlabel{\thesection }}{0pt}
  {\begin{tikzpicture}[remember picture,overlay]
       \end{tikzpicture}
  }
\titlespacing*{\section}{0pt}{0pt}{0pt}

\newcommand*\subsectionlabel{}
\titleformat{\subsection}
  {\gdef\subsectionlabel{}
   \large\bfseries\scshape}
  {\gdef\subsectionlabel{\thesubsection  }}{0pt}
  {\begin{tikzpicture}[remember picture]
    	\draw (-0.15, 0) node[left] {\color{\stylecolor} \textsf{\subsectionlabel}};
	\draw (0.15, 0) node[right] {\color{\stylecolor} \textsf{#1}};
	\fill[color=\stylecolor] (-0.05, -0.23) rectangle (0.05, 0.23);
       \end{tikzpicture}
  }
\titlespacing*{\subsection}{-4pt}{10pt}{0pt}

\newcommand*\subsubsectionlabel{}
\titleformat{\subsubsection}
  {\gdef\subsubsectionlabel{}
   \bfseries\scshape}
  {\gdef\subsubsectionlabel{\thesubsubsection.\ \  }}{0pt}
  {\begin{tikzpicture}[remember picture]
    	\draw (0, 0) node[left] {\color{\stylecolor} \textsf{\subsubsectionlabel}};
	\draw (0, 0) node[right] {\color{\stylecolor} \textsf{#1}};
       \end{tikzpicture}
  }
\titlespacing*{\subsubsection}{-4pt}{7pt}{0pt}

\pgfplotsset{every axis legend/.append style={at={(1.02,1)},anchor=north west}}

\graphicspath{{figures/}}

\begin{document}

\allowdisplaybreaks

\pagestyle{fancy}
\renewcommand{\headrulewidth}{0pt}
\fancyhead{}

\fancyfoot{}
\fancyfoot[C] {\textsf{\textbf{\thepage}}}

\begin{equation*}
\begin{tikzpicture}
\draw (\textwidth, 0) node[text width = \textwidth, right] {\color{white} easter egg};
\end{tikzpicture}
\end{equation*}

\begin{equation*}
\begin{tikzpicture}
\draw (0.5\textwidth, -3) node[text width = \textwidth] {\huge  \textsf{\textbf{Electronic hydrodynamics and the breakdown of the \\ \vspace{0.07in} Wiedemann-Franz and Mott laws in interacting metals}} };
\end{tikzpicture}
\end{equation*}
\begin{equation*}
\begin{tikzpicture}
\draw (0.5\textwidth, 0.1) node[text width=\textwidth] {\large \color{black}  \textsf{Andrew Lucas}$^{\color{\stylecolor} \mathsf{a}}$ \textsf{and  Sankar Das Sarma}$^{\color{\stylecolor} \mathsf{b}}$ };
\draw (0.5\textwidth, -0.5) node[text width=\textwidth] {$^{\color{\stylecolor} \mathsf{a}}$ \small{\textsf{Department of Physics, Stanford University, Stanford, CA 94305, USA}}};
\draw (0.5\textwidth, -0.6) node[text width=\textwidth, below] {$^{\color{\stylecolor} \mathsf{b}}$ {\small \textsf{Condensed Matter Theory Center and Joint Quantum Institute, Department of Physics, University of Maryland, College Park, MD 20742 USA}}};
\end{tikzpicture}
\end{equation*}
\begin{equation*}
\begin{tikzpicture}
\draw (0, -13.1) node[right, text width=0.5\paperwidth] {\texttt{ajlucas@stanford.edu}};
\draw (\textwidth, -13.1) node[left] {\textsf{\today}};
\end{tikzpicture}
\end{equation*}
\begin{equation*}
\begin{tikzpicture}
\draw[very thick, color=\stylecolor] (0.0\textwidth, -5.75) -- (0.99\textwidth, -5.75);
\draw (0.12\textwidth, -6.25) node[left] {\color{\stylecolor}  \textsf{\textbf{Abstract:}}};
\draw (0.53\textwidth, -6) node[below, text width=0.8\textwidth, text justified] {\small We present the theory of thermoelectric transport in metals with long-lived quasiparticles, carefully addressing the interplay of electron-electron scattering as well as electron-impurity scattering, but neglecting electron-phonon scattering.  In Fermi liquids with a large Fermi surface and  weak electron-impurity scattering,   we provide universal and simple formulas for the behavior of the  thermoelectric conductivities across the ballistic-to-hydrodynamic crossover.  In this regime,  the electrical conductivity is relatively unchanged by hydrodynamic effects.  In contrast, the thermal conductivity can be parametrically smaller than predicted by the Wiedemann-Franz law.  A less severe violation of the Mott law arises.   We quantitatively compare the violations of the Wiedemann-Franz law arising from (\emph{i}) momentum-conserving electron-electron scattering in the collision integral, (\emph{ii}) hydrodynamic modifications of the electron-impurity scattering rate, and (\emph{iii}) thermal broadening of the Fermi surface, and show that (\emph{i}) is generally the largest effect.  We present simple formulas for electrical and thermal magnetoconductivity across  the ballistic-to-hydrodynamic limit, along with a more complicated formula for the thermoelectric magnetoconductivity.   In a finite magnetic field, the Lorenz number may be smaller or larger than predicted by the Wiedemann-Franz law, and the crossover between these behaviors is a clear prediction for experiments.  The arbitrarily strong violation of the Wiedemann-Franz law found in our work arises entirely from electron-electron interaction effects within the Fermi liquid paradigm, and does not imply any non-Fermi liquid behavior.  We predict clear experimental signatures of bulk hydrodynamics in high-mobility 2D GaAs semiconductor structures, where a spectacular failure of the Wiedemann-Franz law should persist down to very low temperatures in high-quality and low-density samples.};
\end{tikzpicture}
\end{equation*}

\tableofcontents

\begin{equation*}
\begin{tikzpicture}
\draw[very thick, color=\stylecolor] (0.0\textwidth, -5.75) -- (0.99\textwidth, -5.75);
\end{tikzpicture}
\end{equation*}

\titleformat{\section}
  {\gdef\sectionlabel{}
   \Large\bfseries\scshape}
  {\gdef\sectionlabel{\thesection }}{0pt}
  {\begin{tikzpicture}[remember picture]
	\draw (0.2, 0) node[right] {\color{\stylecolor} \textsf{#1}};
	\draw (0.0, 0) node[left, fill=\stylecolor,minimum height=0.27in, minimum width=0.27in] {\color{white} \textsf{\sectionlabel}};
       \end{tikzpicture}
  }
\titlespacing*{\section}{0pt}{20pt}{5pt}

\section{Introduction}
One of the most important open problems in condensed matter physics is a non-perturbative understanding of interaction effects on the electronic transport properties of metals.  The subject is vast and an enormous literature exists, spanning almost a century.   Due to unresolved experimental puzzles such as the high temperature linear-in-temperature resistivity of the normal phase of high-temperature superconductors at optimal  doping \cite{mackenzie2013}, the study of interaction-limited transport has taken on much significance over the last two decades.    Typically, one assumes that strong electron-electron interactions lead to a breakdown of Fermi liquid physics and of a quasiparticle description, such that the resulting metal is a non-Fermi liquid (NFL) \cite{emery, mahajan, hartnoll1, lucasrmp}.     Despite much effort, the understanding of NFLs is very challenging and remains incomplete.   

 A common heuristic used in experiments to probe the breakdown of Fermi liquid phenomenology is the Wiedemann-Franz (WF) law \cite{wiedemannfranz, lorenz}, which states that the ratio of thermal and electrical conductivity is  \begin{equation}
\frac{\kappa}{\sigma T} \approx \frac{\mpi^2}{3} \frac{k_{\mathrm{B}}^2}{e^2}.  \label{eq:WF}
\end{equation}
The ratio on the left hand side above is called the Lorenz number, and the constant on the right hand side is the ideal Lorenz number experimentally observed in most normal elemental metals at room temperatures.
The WF law arises in a conventional metal at temperatures $T\ll T_{\mathrm{F}}$, the Fermi temperature, whenever electron-electron collisions are negligible, and elastic scattering processes dominate \cite{ziman}.   Transport properties of normal metals are dominated by electron-phonon scattering, which is essentially quasi-elastic at room temperatures.  The theoretical prediction (\ref{eq:WF}) therefore holds very well for most normal metals \cite{prasad}.  Furthermore, NFLs will usually (but not always) manifest a Lorenz number much smaller than predicted by (\ref{eq:WF}) \cite{mahajan}.   In some instances, the implication that bad metals (apparent NFLs with large  resistivity \cite{emery, mahajan, hartnoll1}) violate the WF law has been ``inverted" to conclude that a failure of the WF law by itself is evidence for the existence of a NFL \cite{rwhill}.   Of course, this conclusion is  not logically necessary:  it is possible for Fermi liquids to violate (\ref{eq:WF}) even in the very low temperature limit \cite{vignale}.  In particular, (\ref{eq:WF}) is violated whenever the scattering processes which dominate transport are inelastic.

The purpose of this paper is to -- within the Fermi liquid paradigm -- describe the interplay of electron-electron interactions and impurities.  A careful understanding of this interplay is essential to develop a NFL transport theory.    We will see that some of the thermoelectric transport phenomena that occur in NFLs also occur in FLs.   Our description of `unconventional' FL transport phenomena both provides a ``less strange" setting in which to understand interaction-limited transport physics, and sheds light on which transport phenomena rely on the breakdown of FL theory, and which do not.   Both the hydrodynamic FLs which we consider, and the NFLs outlined previously, are strongly interacting electron systems.  One main difference is that NFL behavior is commonly believed to emerge due to a putative nearby quantum critical point.    The  WF violation and related `exotic' transport properties being studied in our current work arise not from any hidden quantum criticality, but from strong interaction effects within the metallic Fermi liquid itself.  The hidden quantum criticality which may be responsible for NFL physics will most strongly manifest itself in specific temperature-dependent corrections to the effective electron-impurity scattering rate.\footnote{At finite temperatures there is generally simply a  crossover (not a phase transition) between the quantum critical fan and conventional portions of the phase diagram \cite{sachdev}.   The impact of quantum critical fluctuations is most clearly seen in the unconventional temperature dependence of  scattering rates.}  It is an interesting question, beyond the scope of the current work, whether an interplay of quantum criticality and hydrodynamic effects coexist in strongly correlated materials; see recent discussion in \cite{dsz,hartnoll1704}.   

Expanding upon earlier works such as \cite{mahajan, vignale}, we provide a comprehensive and quantitative theory of the finite temperature breakdown of the WF law in metals described by Fermi liquid theory.   In particular,  we confirm the results of \cite{vignale} that the Lorenz ratio becomes parametrically smaller than (\ref{eq:WF}) in the hydrodynamic limit \cite{lucasreview17}, where non-umklapp electron-electron scattering processes occur much faster than electron-imputiy or electron-phonon processes.   This effect is universal and does not rely on any details of the interactions or band structure.   Any metal, if driven to a hydrodynamic  regime,\footnote{In the absence of umklapp, this can be done by making the crystal exceptionally pure so as to drastically reduce the electron-impurity scattering rate.} will necessarily violate (\ref{eq:WF}).  This violation does not  necessarily imply  NFL behavior.   

We will also describe the interaction-driven breakdown of the related Mott law for thermoelectric conductivity:
\begin{equation}
\alpha  = - \frac{\mpi^2}{3}  \frac{k_{\mathrm{B}}^2 T}{e} \frac{\partial \sigma}{\partial \mu},  \label{eq:mott}
\end{equation}
with $\mu = k_{\mathrm{B}}T_{\mathrm{F}}$ the chemical potential, or Fermi energy.   We will see that the Mott  law holds in the same parameter regimes as the WF law; however, its breakdown is not as dramatic.

The reason that hydrodynamic breakdowns of the WF and Mott laws are not observed in typical metals,  even at very low temperatures, is simply a reflection of the fact that most metals are rather dirty, and  exhibit electron-electron scattering rates which are negligibly small compared to electron-impurity scattering rates.  Elemental 3D metals are never in the hydrodynamic regime, even when they are relatively pure -- at low (high) temperatures, electron-impurity (-phonon) interactions are stronger than the corresponding electron-electron interaction.  However, effectively 2D metals (e.g. graphene, high-mobility 2D semiconductor layers) can often be clean enough for the electron-electron scattering rate to surpasss the electron-impurity scattering rate at low to intermediate temperatures (where electron-phonon scattering is relatively weak).   These systems can be in the hydrodynamic regime. We urge experiments in high-mobility low-density modulation-doped 2D GaAs electron and hole systems \cite{dassarmagaas} in order to verify our predictions of hydrodynamics-induced breakdowns of  the WF and Mott relations arising from electron-electron interactions.   These high-quality GaAs systems, which are routinely used for studies of the fractional quantum Hall effect, are exceptionally pure and easily enter the hydrodynamic regime in a 0.1--10 K temperature range.

In fact, recent experiments on graphene \cite{bandurin, levitov1703}, $\mathrm{PdCoO}_2$ \cite{mackenzie}, $\mathrm{WP}_2$ \cite{felser} and GaAs \cite{bakarov} have all observed evidence for the collective hydrodynamic flow of electrons.    Interaction-limited transport  phenomena in low density oxides such as $\mathrm{SrTiO}_3$ \cite{behnia, stemmer} may also have a hydrodynamic character.   Our results complement the existing theoretical literature on hydrodynamic transport \cite{andreev, polini, levitovhydro, lucas3, levitov1607, torre, scaffidi} and contain new hydrodynamic predictions for experiments in the materials listed above.   In graphene, breakdowns of the WF \cite{crossno}  and Mott  \cite{ghahari} laws have been reported and attributed to hydrodynamic effects.  A demonstration of quantitative consistency between  bulk transport phenomena and more direct probes of viscous effects such as nonlocal resistance will be strong evidence for the hydrodynamic nature of electron flow in  these systems.

Consistent with our goal of understanding the role of electron-electron interaction in metallic transport (and particularly, the hydrodynamic regime), we  ignore electron-phonon interaction in our work.  This is not because electron-phonon interaction is generically unimportant -- as we noted  previously,  room temperature transport in normal metals is dominated by phonons -- but because the role of electron-phonon interaction in metallic transport is  well-understood \cite{ziman}.  Neglecting phonons keeps our  theory transparent and tractable.  It is straightforward to add electron-phonon scattering effects to our theory.   

\subsection{Outline}
The rest of the paper is organized as follows.   In Section \ref{sec:main} we summarize some  key results and experimental predictions.   Section \ref{sec:kinetic} outlines the computation of transport coefficients  from kinetic theory.   Detailed calculations of  transport coefficients at low temperature in any Fermi liquid with an isotropic dispersion relation are provided in Section \ref{sec:largeFS} in the absence of a background magnetic field, and in Section \ref{sec:magnetic} in a background magnetic field.  We conclude in Section \ref{sec:conclusion} emphasizing our key results and pointing out experimental implications.  Appendices contain a few technical details of our calculations.

\section{Main Results}
\label{sec:main}
In this paper, we compute the dc thermoelectric transport coefficients of a metal:   \begin{equation}
\left(\begin{array}{c}  J_i \\ Q_i\end{array}\right) = \left(\begin{array}{cc}  \sigma_{ij}  &\ T\alpha_{ij} \\ T\alpha_{ij} &\  T\bar\kappa_{ij} \end{array}\right)  \left(\begin{array}{c}  E_j \\ - \frac{1}{T} \partial_j T \end{array}\right),  \label{eq:transdef}
\end{equation}
where $J_i$ is the charge current, $Q_i$ is the heat current, $E_j$ is an externally applied electric field, and $\partial_j T$ is an ``externally applied temperature gradient".\footnote{A formal discussion of how this can be done may be found in \cite{lucasrmp}.}   In experiments, one often measures the open circuit thermal  conductivity, where no electrical current flows in the system:
 \begin{equation}
\kappa_{ij} = \bar\kappa_{ij} - T\alpha_{ik} \sigma^{-1}_{kl} \alpha_{lj}.
\end{equation}
For much of this paper, we will assume that the conductivity tensors are isotropic:  e.g. $\sigma_{ij} = \sigma\mdelta_{ij}$.

 A number of systems ranging from graphene \cite{crossno, ghahari} to more exotic ``strange metals" \cite{rwhill, rpsmith, pfau} violate the relations (\ref{eq:WF}) and (\ref{eq:mott}).   Some of these violations could arise from the dominance of inelastic scattering processes, while others appear to have a hydrodynamic origin \cite{crossno}.   The purpose of this paper will be to elucidate and expand upon the breakdown of (\ref{eq:WF}) and (\ref{eq:mott}) due to hydrodynamic effects.  Indeed, it is well-understood why in the hydrodynamic regime (\ref{eq:WF})) and (\ref{eq:mott}) will fail.   For example, a  new universal transport relation arises deep in the hydrodynamic regime: \cite{hkms, hofman,  lucasMM}  \begin{equation}
\sigma = \frac{e^2 n^2}{Ts^2} \bar\kappa = \frac{-en}{s} \alpha.  \label{eq:introhydro}
\end{equation}
Here $s$ is the entropy density, $n$ is the number density of electrons, and $-e$ is the charge of the electron.   These ratios emerge because all thermoelectric transport phenomena become linked to momentum-relaxing scattering processes.    Also observe that in this hydrodynamic regime, $\kappa \ll \sigma$ because $\bar\kappa \sigma \approx T\alpha^2$ \cite{mahajan};  furthermore, $\bar\kappa \gg \kappa$. 

The first main result of this paper is the quantitatively accurate description of the transition between the conventional ``collisionless" regime (\ref{eq:WF}) and (\ref{eq:mott}), and the hydrodynamic regime (\ref{eq:introhydro}), carefully accounting for all hydrodynamic effects.   In metals with large Fermi surfaces and  weak disorder, we will show that \begin{equation}
\frac{\kappa}{\sigma T} \approx \frac{\mpi^2}{3} \frac{\Gamma}{\Gamma+\gamma}, \label{eq:33WF}
\end{equation}
where $\gamma$  denotes the electron-electron scattering rate, and $\Gamma$ denotes the electron-impurity scattering rate.   While (\ref{eq:33WF}) was derived previously in \cite{vignale}, our derivation will naturally generalize to the Mott law, and to more complicated settings.   We will describe the computation of $\Gamma$ in some detail below,  including a qualitative discussion of hydrodynamic effects on $\Gamma$ \cite{hartnoll1706, lucasRFB} in Section \ref{sec:eeimp}.   We briefly discuss finite-temperature  corrections to (\ref{eq:33WF}) in $T/T_{\mathrm{F}}$ in Section \ref{sec:weakint}.  Eq. (\ref{eq:33WF}) makes clear that as observed prior,  $\kappa/\sigma T$ is generally smaller than the Wiedemann-Franz prediction in an interacting metal at finite density.    Eq. (\ref{eq:33WF}) also implies the expected FL behavior at $T=0$, where $\gamma=0$.  This is in contrast to a NFL, where it is possible for the WF relation to be violated even at $T=0$.    It would be interesting to experimentally study the behavior of the Lorenz number at very low temperatures in strange metals (although in some cases, the putative critical point is unstable to superconductivity, e.g. cuprates, making a low temperature measurement impractical).  The recovery of the WF law at low temperatures, or lack  thereof,  is a simple check on the FL paradigm.


The second main result  of this paper is the generalization of (\ref{eq:33WF}) to magnetotransport. Including a background magnetic field which leads to cyclotron frequency $\omega_{\mathrm{c}}$,  we find that when $T\ll T_{\mathrm{F}}$: \begin{subequations}\begin{align}
\frac{\kappa_{xx}}{\sigma_{xx} T} &\approx \frac{\mpi^2}{3e^2} \frac{\Gamma+\gamma}{\Gamma}  \frac{\Gamma^2+\omega_{\mathrm{c}}^2}{(\Gamma+\gamma)^2+\omega_{\mathrm{c}}^2} ,  \label{eq:sec2main} \\
\frac{\kappa_{xy}}{\sigma_{xy} T} &\approx \frac{\mpi^2}{3e^2} \frac{\Gamma^2+\omega_{\mathrm{c}}^2}{(\Gamma+\gamma)^2+\omega_{\mathrm{c}}^2} .
\end{align}\end{subequations}
Like (\ref{eq:33WF}), the ratio of the Hall thermal to Hall electrical conductivity \emph{strictly decreases} with increasing electron-electron interactions.    However, the dissipative conductivities exhibit a richer structure.   When $\Gamma+\gamma < \omega_{\mathrm{c}}$,  enhancing electron-electron interactions increases the Lorenz number \emph{above} the ballistic prediction;  only when $\Gamma+\gamma > \omega_{\mathrm{c}}$ does the Lorenz number become smaller than (\ref{eq:WF}).   This implies that a non-monotonic temperature dependence of the Lorenz number will occur whenever $\Gamma<\omega_{\mathrm{c}}$.   It is possible, though  difficult, to accurately measure electronic thermal conductivity directly \cite{crossno, fong, crossno2}.   Eq. (\ref{eq:sec2main}) is a key prediction for future experiments and will provide a quantitative test of both the hydrodynamic origin of Wiedemann-Franz violations, and  the validity of the Fermi liquid paradigm.

\section{Kinetic Theory Formalism}
\label{sec:kinetic}
In this section, we introduce the general formalism  to solve the transport problem.   We will start out with rather minimal assumptions beyond the existence of quasiparticles with lifetime $\gg \hbar /k_{\mathrm{B}}T$, and as the section  continues we will make more and more specific assumptions.    It  is (at least in principle) simple to  relax many of these assumptions.   In practice, going beyond some of these assumptions may necessitate extensive numerical work focusing on specific materials, which is beyond the scope of our current work.\footnote{Also, the detailed system parameters for carrying out such numerical calculations for specific strongly correlated materials may not be known.  For such systems it would not yet be useful to perform serious numerical computations.}

\subsection{Linearized Boltzmann  Equation}
In order to compute the thermoelectric conductivity matrix, we solve the Boltzmann equation for a Fermi liquid of dispersion relation $\epsilon(\mathbf{p})$, linearized about thermal equilibrium at temperature $T$.    We assume that the band structure is inversion-symmetric and time-reversal invariant.   We also assume the presence of an inhomogeneous single-particle potential energy  $V_{\mathrm{imp}}(\mathbf{x})$ caused by random impurities.     We ignore the spin of  the electronic quasiparticles, as we are explicitly interested in spin-independent transport properties, though it is easy to keep track of spin if necessary.

We denote by $f(\mathbf{x},\mathbf{p})$ the distribution function of (quantum) kinetic theory.   Letting $f=f_{\mathrm{eq}} + \mdelta f$, where \begin{equation}
f_{\mathrm{eq}}(\mathbf{x},\mathbf{p}) = \frac{1}{1+\mathrm{e}^{(\epsilon(\mathbf{p}) + V_{\mathrm{imp}}(\mathbf{x})-\mu)/k_{\mathrm{B}}T}},
\end{equation}
denote  the distribution function in the absence of any external electric fields or temperature gradients, and $\mdelta f$  denote the infinitesimal correction arising in infinitesimal electric fields/temperature gradients,
we find that the Boltzmann equation reduces to \begin{equation}
\partial_t \mdelta f  + \mathbf{v} \cdot \frac{\partial \mdelta f}{\partial \mathbf{x}}+ \mathbf{F}_{\mathrm{imp}} \cdot \frac{\partial \mdelta f}{\partial \mathbf{p}} +  \mdelta \mathbf{F}_{\mathrm{ext}} \cdot \frac{\partial f_{\mathrm{eq}}}{\partial \mathbf{p}} = -\mdelta \mathcal{C}[\mdelta f]  \label{eq:linboltz}
\end{equation}
within linear response.   As we are only interested in the dc transport problem, we may drop the time  derivative from now on. Here \begin{equation}
\mathbf{v} = \frac{\partial \epsilon}{\partial \mathbf{p}}
\end{equation} 
$\mathbf{F}_{\mathrm{imp}} = -\nabla V_{\mathrm{imp}}$ are the forces arising from the  impurity potential,  and \begin{equation}
\mdelta \mathbf{F}_{\mathrm{ext}} = -e\left(\mdelta \mathbf{E} - \frac{\epsilon - \mu}{T}\nabla \mdelta T\right).
\end{equation}
$\mdelta \mathcal{C}[\mdelta f]$ denotes the collision  integral, written only to leading (first) order in $\mdelta f$.   Explicit discussions of the linearized collision  integral may be found in \cite{ledwith1}.    A solution of the linearized Boltzmann equation (\ref{eq:linboltz}) will provide $\mdelta  f$.   As there are no charge and heat currents in equilibrium,  the charge and heat currents are given entirely by this perturbation:\begin{subequations}\begin{align}
J_i(\mathbf{x}) &=  -e \int \frac{\mathrm{d}^d\mathbf{p}}{(2\mpi \hbar)^d} v_i \mdelta f(\mathbf{x},\mathbf{p}), \\
Q_i(\mathbf{x}) &=   \int \frac{\mathrm{d}^d\mathbf{p}}{(2\mpi \hbar)^d} (\epsilon + V_{\mathrm{imp}}-\mu) v_i \mdelta f(\mathbf{x},\mathbf{p}),
\end{align}\end{subequations}
For  simplicity in this paper, and without loss of generality, we will shift $\epsilon(\mathbf{p})$  by a constant such that the Fermi surface is located at $\mu=0$, which simplifies a few of the formulas below.

Mathematically, the Boltzmann equation is simply a linear algebra problem in a high-dimensional  vector space.  In order to make analytic progress, we cannot simply invert  the linearized Boltzmann operator --  it is far too complicated.  We instead proceed in a  number of steps, which may (for the moment) appear rather cumbersome, but will in the end reduce the calculation of the conductivities to the inversion of  few-dimensional matrices,  in certain limits.   The notation and formulations below follow closely \cite{hartnoll1706, lucasRFB}.

The  first step is  to  define a suitable inner product for the  $\mathbf{p}$-indices on our vector space: \begin{equation}
\langle  f|g\rangle  = \int  \frac{\mathrm{d}^d\mathbf{p}}{(2\mpi\hbar)^d}  \left(-\frac{\partial f_{\mathrm{eq}}}{\partial \epsilon}\right) f(\mathbf{p})g(\mathbf{p}),  \label{eq:innerproduct}
\end{equation}
Writing the distribution function as  \begin{equation}
f(\mathbf{x},\mathbf{p}) = f_{\mathrm{eq}}(\mathbf{x},\mathbf{p}) + \int \frac{\mathrm{d}^d\mathbf{k}}{(2\mpi)^d}\left(-\frac{\partial f_{\mathrm{eq}}}{\partial \epsilon}\right) \Phi(\mathbf{x},\mathbf{p}),
\end{equation}
with $\Phi$ treated as infinitesimally small, and defining  \begin{subequations}\begin{align}
|\mathsf{J}_i\rangle &=  \int  \mathrm{d}^d\mathbf{p} \;    v_i(\mathbf{p}) | \mathbf{p}\rangle , \\
|\mathsf{Q}_i\rangle &=  \int  \mathrm{d}^d\mathbf{p} \;   \epsilon(\mathbf{p})   v_i(\mathbf{p}) | \mathbf{p}\rangle,  \\
\end{align}\end{subequations}
we find that \begin{subequations}\begin{align}
\langle \Phi(\mathbf{x})| \mathsf{J}_i  \rangle &= J_i (\mathbf{x}), \\
\langle \Phi(\mathbf{x})| \mathsf{Q}_i  \rangle &= Q_i (\mathbf{x}) - V_{\mathrm{imp}}(\mathbf{x})J_i(\mathbf{x}).
\end{align}\end{subequations}
The advantage of introducing the inner product (\ref{eq:innerproduct}) is now clear:  the highly singular structure of the Fermi function has been absorbed into the integration measure in the inner product.  The vectors which we are studying are now smooth functions of momentum across  the Fermi surface.   We also emphasize that the $V_{\mathrm{imp}}$-dependent correction  to the heat current will be negligible in the limits studied in this paper.   We will think of $|\mathsf{J}_i\rangle$ as the vector that encodes the charge current, and $|\mathsf{Q}_i\rangle$ as the vector that encodes the heat current.

Let  $\mathsf{W}$ denote the linearized collision integral.  It is a symmetric, positive-semidefinite matrix with the assumptions made above;  its null vectors are associated with local conservation laws of electron-electron  collisions.   We will  assume  that one of these conservation laws  is associated with momentum, because translation invariance is only broken by $V_{\mathrm{imp}}$.  The linearized Boltzmann equation (\ref{eq:linboltz}) can then be written in  our vector notation as \begin{equation}
 \langle \mathbf{p}| \left[\mathbf{v}\cdot  \frac{\partial}{\partial \mathbf{x}} + \mathbf{F}_{\mathrm{imp}}\cdot \frac{\partial}{\partial \mathbf{p}}\right]|\Phi\rangle   = -    \langle \mathbf{p} | \mathsf{W}| \Phi\rangle + E_i \langle \mathbf{p}| \mathsf{J}_i\rangle - \frac{\partial_i T}{T}  ( \langle \mathbf{p}|\mathsf{Q}_i\rangle +  V_{\mathrm{imp}} \langle \mathbf{p}|\mathsf{J}_i\rangle ) .
\end{equation}
The $(\mathbf{x},\mathbf{p})$-dependence in the inner product has been properly taken into account -- the derivatives above only act on $\Phi$.     Clearly, we may undo the inner product with $|\mathbf{p}\rangle$, since the equation is valid for all $|\mathbf{p}\rangle$.

\subsection{Perturbative Limit}
We now focus on a perturbative limit where the amplitude of the smooth impurity potential $V_{\mathrm{imp}}$ is  perturbatively weak.    This is a natural limit of interest for the study of hydrodynamic interaction effects, and helps ensure that electron-electron interaction effects are stronger than electron-impurity interactions.  Note that although $V_{\mathrm{imp}}$ is small, it cannot vanish because then strict translational invariance is restored and the electrical conductivity is infinite,  in the absence of umklapp (which is ignored throughout this work).   This can be seen by writing down Newton's  second law:  \begin{equation}
\frac{\mathrm{d}\mathbf{P}_{\mathrm{tot}}}{\mathrm{d}t} = Q_{\mathrm{tot}} \mathbf{E}  \label{eq:newton2law}
\end{equation}
where $\mathbf{P}_{\mathrm{tot}}$ is  the total momentum in the theory, and $Q_{\mathrm{tot}}$ is the total charge.    In the absence of translation symmetry breaking, no further terms may be written down on the right hand side, and transport properties in the strict dc limit are ill-posed.
In a quantum theory, (\ref{eq:newton2law}) continues to hold:  it is the Ward identity associated with translation symmetry.  Therefore, we work consistently in the limit of small, but finite, $V_{\mathrm{imp}}$.    

As is well-known, and we will see explicitly below,  all thermoelectric conductivities now scale like $V_{\mathrm{imp}}^{-2}$ as a consequence  of the slowness of momentum relaxation \cite{lucasrmp, lucasreview17, hofman, lucasMM}.    This means that most -- but not all --  $V_{\mathrm{imp}}$-dependence in the problem can  be ignored.   To see where we can neglect $V_{\mathrm{imp}}$, recall that our goal is to  compute \begin{subequations}\begin{align}
J_i^{\mathrm{avg}} &= \int \frac{\mathrm{d}^d\mathbf{x}}{V_d}  \; J_i(\mathbf{x}) \approx \int \frac{\mathrm{d}^d\mathbf{x}}{V_d} \langle  \Phi(\mathbf{x})|\mathsf{J}_i\rangle  \\
Q_i^{\mathrm{avg}} &= \int \frac{\mathrm{d}^d\mathbf{x}}{V_d}  \; Q_i(\mathbf{x}) \approx \int \frac{\mathrm{d}^d\mathbf{x}}{V_d} \langle  \Phi(\mathbf{x})|\mathsf{Q}_i\rangle. 
\end{align}\end{subequations}
Here $V_d$ denotes the spatial  volume -- we  are simply taking a spatial average.   The  latter approximations above come from the fact that both $\sigma$ and $\kappa$ will scale as $V_{\mathrm{imp}}^{-2}$ at leading order.    It  will be useful  to Fourier transform $\Phi$:  \begin{equation}
|\Phi(\mathbf{k})\rangle  \equiv \int \frac{\mathrm{d}^d\mathbf{x}}{V_d}  |\Phi(\mathbf{x}\rangle \mathrm{e}^{-\mathrm{i}\mathbf{k}\cdot\mathbf{x}}.
\end{equation}
(\ref{eq:linboltz}) can be written as
 \begin{equation}
  \left(\begin{array}{cc}  \mathsf{W} &\  \mathsf{L}^0_{\mathrm{dis}} \\ -{\mathsf{L}^0_{\mathrm{dis}}}^\dagger &\   \mathsf{W} + \mathrm{i}\mathbf{k}\cdot \mathbf{v} + \mathsf{L}_{\mathrm{dis}}(\mathbf{k},\mathbf{k}^\prime)  \end{array}\right)    \left(\begin{array}{c}   |\Phi(\mathbf{k}=\mathbf{0})\rangle  \\ |\Phi(\mathbf{k}\ne\mathbf{0})\rangle  \end{array}\right)  = \left(\begin{array}{c}  E_i |\mathsf{J}_i\rangle - \frac{\partial_i T}{T} |\mathsf{Q}_i\rangle  \\ 0  \end{array}\right),   \label{eq:22mat}
\end{equation}
where \begin{equation}
\mathsf{L}_{\mathrm{dis}}(\mathbf{k},\mathbf{k}^\prime) = -\mathrm{i}(k-k^\prime)_i V_{\mathrm{imp}}(\mathbf{k}-\mathbf{k}^\prime) \overrightarrow{\frac{\partial}{\partial p_i}},
\end{equation}
and we have denoted $\mathsf{L}^0_{\mathrm{dis}} = \mathsf{L}_{\mathrm{dis}}(\mathbf{0},\mathbf{k})$.   
Since $V_{\mathrm{imp}}$ is  perturbatively small, and $\mathrm{i}\mathbf{k}\cdot \mathbf{v} + \mathsf{W}$ has no null vectors in general,  we do not need to worry about the $\mathsf{L}_{\mathrm{dis}}$ except in the off-diagonal blocks in (\ref{eq:22mat}).

Combining all of the above together, we conclude that the thermoelectric conductivity matrix is given by \begin{equation}
\left(\begin{array}{cc}   \sigma_{ij} &\ T\alpha_{ij} \\ T\alpha_{ij} &\ T\bar\kappa_{ij} \end{array}\right) \approx \left(\begin{array}{cc}   \langle \mathsf{J}_i| \widetilde{\mathsf{W}}^{-1} |\mathsf{J}_j\rangle &\ \langle \mathsf{J}_i| \widetilde{\mathsf{W}}^{-1} |\mathsf{Q}_j\rangle \\ \langle \mathsf{Q}_i| \widetilde{\mathsf{W}}^{-1} |\mathsf{J}_j\rangle &\ \langle \mathsf{Q}_i| \widetilde{\mathsf{W}}^{-1} |\mathsf{Q}_j\rangle\end{array}\right),  \label{eq:thermoelcondboltz}
\end{equation}
where we have defined the matrix \begin{equation}
\widetilde{\mathsf{W}} \approx \mathsf{W} + \int \frac{\mathrm{d}^d\mathbf{k}}{(2\mpi)^d}  k_i k_j |V_{\mathrm{imp}}(\mathbf{k})|^2  \overleftarrow{\frac{\partial}{\partial  p_i}} (\mathsf{W}+\mathrm{i}\mathbf{k}\cdot  \mathbf{v})^{-1} \overrightarrow{\frac{\partial}{\partial  p_j}} +  \mathrm{O}(V_{\mathrm{imp}}^3).  \label{eq:Wtildedef}
\end{equation}
The first  term  in $\widetilde{\mathsf{W}}$  corresponds to electron-electron interactions.   The second term corresponds to electron-impurity collisions, which will not conserve momentum.   Formally speaking, we observe that the presence of $\mathsf{W}$ in the electron-impurity term means  that the electron-electron interactions can substantially modify the nature  of electron-impurity scattering.  This interplay of electron-electron and electron-impurity interactions is described in some detail in \cite{hartnoll1706, lucasRFB}, and in Section \ref{sec:eeimp}.   However, for  most of this paper, we will focus  on a  limit where $\mathsf{W}$ is \emph{also}  taken to be  perturbatively small:    $\mathsf{W} \sim V_{\mathrm{imp}}^2$.   In this limit, impurity scattering is well described by the single-particle theory: we may approximate \begin{equation}
\widetilde{\mathsf{W}} \approx \mathsf{W} +  \mathsf{W}_{\mathrm{imp}},  \label{eq:Wsum}
\end{equation}
where \begin{equation}
\mathsf{W}_{\mathrm{imp}} = \int \frac{\mathrm{d}^d\mathbf{k}}{(2\mpi)^d}  k_i k_j |V_{\mathrm{imp}}(\mathbf{k})|^2  \overleftarrow{\frac{\partial}{\partial  p_i}}  \mdelta (k_i v_i) \overrightarrow{\frac{\partial}{\partial  p_j}}.  \label{eq:Wimp1}
\end{equation}
Eq. (\ref{eq:Wsum}) is reminiscent of Mattheisen's rule -- the impurity-scattering and electron-scattering diagrams may be added independently to the (disorder-averaged) collision integral of  kinetic theory.   However, we emphasize that in the  final  transport coefficients, Mattheisen's rule for resistivities can be \emph{violated.}    

%

\subsection{Rotational Invariance and a Basis}
The  final assumption that we will make in this paper is that the dispersion relation (and disorder distribution) are rotationally invariant.     This is a  reasonable assumption for many correlated Fermi liquids where the hydrodynamic regime is  likely experimentally accessible,  including graphene and GaAs.     We further assume that $\mathsf{W}$ is  rotationally invariant;   by construction in (\ref{eq:Wimp1}),  it is clear that (after performing the $\mathbf{k}$-integral) $\mathsf{W}_{\mathrm{imp}}$ is rotationally invariant.   Thus,  $\widetilde{\mathsf{W}}$ is  rotationally invariant.  What this means is that the natural basis  of vectors $|\mathbf{p}\rangle$ are:  \begin{equation}
|\alpha,m_1\cdots m_{d-1}\rangle  = \int \mathrm{d}^d\mathbf{p} \;   f_\alpha(p) \mathrm{Y}_{m_1\cdots m_{d-1}}(\theta_1,\ldots,\theta_{d-1}) |\mathbf{p}\rangle
\end{equation}
In the above equation, and henceforth, $p=|\mathbf{p}|$, $\theta_1,\ldots,\theta_{d-1}$ denote the angular coordinates of $\mathbf{p}$, and $f_\alpha$ is a set of polynomial functions that we will detail  below.   In this basis, \begin{equation}
\langle \alpha, m_1\cdots m_{d-1}| \widetilde{\mathsf{W}}|\beta, m_1^\prime \cdots m^\prime_{d-1}\rangle = W_{\alpha\beta}^{m_1\cdots m_{d-1}} \mdelta_{m_1m_1^\prime}\times \cdots \times \mdelta_{m_{d-1}m_{d-1}^\prime}.
\end{equation}

Our goal is to compute the thermoelectric conductivity matrix.   Both $|\mathsf{J}_i\rangle$ and $|\mathsf{Q}_i\rangle$ are vectors  under spatial rotation, and without loss of generality, we may compute $\sigma = \langle \mathsf{J}_x|\widetilde{\mathsf{W}}^{-1}|\mathsf{J}_x\rangle$, etc.   So a natural choice of basis vectors to study corresponds to \begin{equation}
|\alpha\rangle \equiv \frac{p_x}{p}  f_\alpha(p), \label{eq:alphabasis}
\end{equation}
for a basis of polynomials $\lbrace f_\alpha\rbrace$.    For these basis vectors, we observe that \begin{equation}
\langle \alpha | \beta \rangle = \int \frac{\mathrm{d}^d\mathbf{p}}{(2\mpi\hbar)^d} f_\alpha(p) f_\beta(p) \frac{p_x^2}{p^2} = \frac{1}{d} \int \frac{\mathrm{d}^d\mathbf{p}}{(2\mpi\hbar)^d} f_\alpha(p) f_\beta(p). \label{eq:dfactor}
\end{equation}
Up to the overall prefactor of $1/d$, we may simply compute the radial integrals over the Fermi surface.

Another useful simplification in this basis arises when computing the matrix elements $\langle \alpha | \mathsf{W}_{\mathrm{imp}} | \beta\rangle$.    Using  \begin{equation}
k_i \frac{\partial}{\partial p_i} \left(\frac{p_x}{p} f_\alpha \right) = \frac{k_x}{p} f_\alpha  +\left(f_\alpha^\prime -1\right) \frac{k_i p_i p_x}{p^2} ,
\end{equation}
and observing that the $\mdelta(k_i v_i)$ in (\ref{eq:Wimp1}) constrains $k_i p_i = 0$ in an isotropic theory, we conclude  that \begin{align} \label{eq:alphaWimpbeta}
\langle \alpha | \mathsf{W}_{\mathrm{imp}} |\beta\rangle  &= \Omega_d \int  \frac{p^{d-1}\mathrm{d}p}{(2\mpi\hbar)^d} \left(-\frac{\partial f_{\mathrm{eq}}}{\partial \epsilon}\right)   \int \frac{\mathrm{d}^d\mathbf{k}}{(2\mpi)^d}  k_x^2 |V_{\mathrm{imp}}(\mathbf{k})|^2   \mdelta (k_i v_i)  \frac{f_\alpha f_\beta}{p^2} \notag \\
&= \Omega_d \int  \frac{p^{d-1}\mathrm{d}p}{(2\mpi\hbar)^d} \left(-\frac{\partial f_{\mathrm{eq}}}{\partial \epsilon}\right) \frac{f_\alpha f_\beta}{p^2 v}  \left[ \int \frac{\mathrm{d}^d\mathbf{k}}{(2\mpi)^d}  \frac{k^2}{d} |V_{\mathrm{imp}}(\mathbf{k})|^2   \mdelta (k_x) \right] \notag  \\
&=   \mathcal{A}_{\mathrm{imp}} \times \Omega_d \int  \frac{p^{d-1}\mathrm{d}p}{(2\mpi\hbar)^d} \left(-\frac{\partial f_{\mathrm{eq}}}{\partial \epsilon}\right) \frac{f_\alpha f_\beta}{p^2 v},
\end{align}
where $\Omega_d$ is the area of the unit sphere.   Hence, all of the physics of  impurity scattering  becomes captured by a single constant $\mathcal{A}_{\mathrm{imp}}$, which is defined as the object in square brackets in the second line of (\ref{eq:alphaWimpbeta}).  We caution that there can  be non-trivial  temperature dependence in $\mathcal{A}_{\mathrm{imp}}$, in the presence  of temperature-dependent screening of a long-range Coulomb impurity potential $V_{\mathrm{imp}}(\mathbf{k})$ \cite{dassarma1999, dassarma2004}, an effect that we ignore in the present work.

\section{Large Fermi Surfaces}\label{sec:largeFS}
To make further progress, we will now work  in the limit where the Fermi surface is large.   To  be precise, we will assume that $T \ll p_{\mathrm{F}} v_{\mathrm{F}}$.   In this limit, we can further simplify the evaluation of the conductivity by neglecting all but a few  low order polynomials $f_\alpha$  in the basis (\ref{eq:alphabasis}).
\subsection{An Efficient Basis}
\label{sec:basis}
Let   the dispersion relation be \begin{equation}
\epsilon(p) = v_{\mathrm{F}}q + \frac{a}{2}q^2 + \frac{b}{3}q^3 + \cdots,  \label{eq:disp}
\end{equation}
A natural set of basis functions can be constructed as follows.  Letting $q\equiv p-p_{\mathrm{F}}$, we define the states $|q^n\rangle$,  corresponding to the polynomials $f_\alpha =  q^n$.   Of great importance to us is that in this basis,
\begin{subequations}\label{eq:JxQx31}\begin{align}
|\mathsf{J}_x\rangle &= -e\left(v_{\mathrm{F}}|q^0\rangle + a|q^1\rangle + b|q^2\rangle +  \cdots\right), \\
|\mathsf{Q}_x\rangle &= v_{\mathrm{F}}^2|q^1\rangle + \frac{3}{2}v_{\mathrm{F}}a|q^2\rangle  +  \cdots.
\end{align}\end{subequations}

Unfortunately, the basis functions $|q^n\rangle$ are not normalized.   It would  be more convenient if we could work  with an orthonormal basis.  This can be constructed straightforwardly by the Gram-Schmidt procedure, and we will denote the result of this procedure  with $|n\rangle$.   The calculation of $|n\rangle$ from $|q^n\rangle$  is  straightforward  but tedious:  details are provided in Appendix \ref{app:details}.   A useful qualitative fact for the discussion that follows is that \begin{equation}
\langle q^n|q^m\rangle \sim \left\lbrace \begin{array}{ll} T^{n+m} &\ n+m \text{ even} \\ T^{n+m+1} &\ n+m \text{ odd} \end{array}\right..
\end{equation}  In  what follows, \emph{subleading  numerical coefficients in $T$ are  specific to $d=2$ spatial dimensions},  but the  general structure is unchanged for all dimensions.    The first  few basis  functions are \begin{subequations}\label{eq:basischange}\begin{align}
|0\rangle &=  c_{00} \sqrt{\frac{d}{\nu}} |q^0\rangle , \\
|1\rangle &= \sqrt{\frac{d}{\nu}}\left[  \frac{c_{11}}{T}  |q^1\rangle + Tc_{10} |q^0\rangle\right],
\end{align}\end{subequations}
with $\nu$  the density of states (we remind the reader of useful thermodynamic formulas for a Fermi liquid in the low temperature limit in Appendix \ref{app:thermo}), the factor of $d$ arising from (\ref{eq:dfactor}), and
\begin{subequations}\begin{align}
c_{00} &= 1 - \frac{\mpi^2T^2}{2}\left(\frac{a^2}{2v_{\mathrm{F}}^4} - \frac{a}{2p_{\mathrm{F}}v_{\mathrm{F}}^3} - \frac{b}{3v_{\mathrm{F}}^3}\right) + \mathrm{O}(T^4), \\
c_{11} &= \frac{\sqrt{3}v_{\mathrm{F}}}{\mpi} + \frac{\mpi T^2  \left(2v_{\mathrm{F}}^2 + 14 bv_{\mathrm{F}}p_{\mathrm{F}}^2 + 15ap_{\mathrm{F}}v_{\mathrm{F}} - 27a^2p_{\mathrm{F}}^2 \right)}{4\sqrt{3}v_{\mathrm{F}}^3p_{\mathrm{F}}^2} + \mathrm{O}(T^4) , \\
c_{10} &= \frac{\mpi (3ap_{\mathrm{F}}-2v_{\mathrm{F}})}{2\sqrt{3}p_{\mathrm{F}}v_{\mathrm{F}}^2} + \mathrm{O}(T^2).
\end{align}\end{subequations}
A general basis function is \begin{equation}
|n\rangle =  \sqrt{\frac{d}{\nu}} \left(\frac{v_{\mathrm{F}}}{T}\right)^n c_{nn} |q^n\rangle + \left(\frac{v_{\mathrm{F}}}{T}\right)^{n-2} \left(c_{n,n-1}|q^{n-1}\rangle + c_{n,n-2}|q^{n-2}\rangle\right)  + \cdots.   \label{eq:basisexpansion}
\end{equation}
A key observation is that -- up to $\mathrm{O}(T^2/T_{\mathrm{F}}^2)$ corrections  -- the  orthonormal basis is equivalent to $|q^n\rangle$, up to a prefactor which is propotional to $T^{-n}$.   Also observe that $|q^n\rangle$ can \emph{exactly} be expressed as a function of $|m\rangle$ with $m\le n$.

\subsection{Approximating the Collision Integrals}

From the form of (\ref{eq:JxQx31}) and (\ref{eq:basischange}), it is clear that as long as components of $\mathsf{W}_{\mathrm{imp}}$ are ``well-behaved" in the low $T$ limit, then we can efficiently compute the thermoelectric conductivity matrix by focusing only on the first few basis vectors:  $|0\rangle, |1\rangle, \ldots$.    Indeed, this is the case: 
\begin{equation}
 \langle m|\mathsf{W}_{\mathrm{imp}}|n\rangle \lesssim T^{|m-n|}.
 \end{equation}
 To show this, we first write
  \begin{equation}
\frac{\mathcal{A}_{\mathrm{imp}}}{p^2v(p)} = \Gamma + \Gamma_1 q + \Gamma_2 q^2 + \cdots.  \label{eq:Gammadef}
\end{equation}
By definition,  $q^k|q^n\rangle  = |q^{n+k}\rangle$.   (We are denoting the linear operation which multiplies by $q$ as $q$, in the obvious way.)   Now, assuming that $m>n$, without loss of  generality, observe that \begin{equation}
\langle m| q^{m-n}|n\rangle \sim \langle m| \left[ \frac{1}{T^n} |q^m\rangle + \frac{T^2}{T^n} |q^{m-1}\rangle + \frac{T^2}{T^n} |q^{m-2}\rangle + \cdots  \right] \sim \frac{\langle m|q^m\rangle}{T^n} \sim T^{m-n}.
\end{equation}
In fact, staring at the above expression,  we also find that $\langle m-\ell |q^{m-n}|q^n\rangle \lesssim T^{m-n}$ for any $\ell\ge 0$.
Therefore, \begin{equation}
\langle m| \mathsf{W}_{\mathrm{imp}}|n\rangle \sim \Gamma_{m-n} T^{m-n} + \cdots. \label{eq:Gammamn}
\end{equation}
The $\cdots$ in (\ref{eq:Gammamn}) includes terms  proportional to $\Gamma_{m-n+2}T^{m-n+2}$, etc., which are all subleading in the low temperature limit.   At $T=0$, there is a residual resistivity: \begin{equation}
\langle m| \mathsf{W}_{\mathrm{imp}}|n\rangle = \Gamma \mdelta_{mn}.  \label{eq:T0residual}
\end{equation}
We have not dropped any constant prefactor in the above formula (\ref{eq:T0residual}).

We next observe that \begin{equation}
\langle m| \mathsf{W}_{\mathrm{imp}}^{-1}|n\rangle \lesssim T^{|m-n|}.
\end{equation}
One way to  show this result is recursively.  From the form  of the $T=0$ result (\ref{eq:T0residual}), we show the base case $|n-m|=0$.   For finite $|n-m|$, we  may use block matrix inversion identities.   Without loss of generality,  we again take $m>n$.   Let us write \begin{equation}
\mathsf{W} = \left(\begin{array}{cc}  \langle  k < m| &\ \langle k\ge m| \end{array}\right) \left(\begin{array}{cc}  \mathsf{W}_{--} &\ \mathsf{W}_{-+}  \\ \mathsf{W}_{+-} &\ \mathsf{W}_{++} \end{array}\right)\left(\begin{array}{c}  |  k < m\rangle  \\ | k\ge m\rangle \end{array}\right)
\end{equation}  Since  there are no singular eigenvalues at low temperatures, \begin{equation}
\langle m| \mathsf{W}_{\mathrm{imp}}^{-1}|n\rangle \sim \sum_{k<m, \ell \ge m} \left(\mathsf{W}_{++} - \mathsf{W}_{+-}\mathsf{W}_{--}^{-1}\mathsf{W}_{-+}\right)^{-1}_{m\ell} (\mathsf{W}_{+-})_{\ell k} (\mathsf{W}_{--}^{-1})_{kn} \lesssim \sum_{k<m, \ell \ge m} T^{|k-\ell | + |k-n|} \lesssim T^{|m-n|} .
\end{equation}


Since physical observables such as $|\mathsf{J}_x\rangle$ and $|\mathsf{Q}_x\rangle$ are naturally expressed in terms of the $|q^n\rangle$ basis, as in (\ref{eq:JxQx31}), we are  only interested in evaluating $\mathsf{W}_{\mathrm{imp}}^{-1}$ on  vectors  which take the form \begin{equation}
|\mathsf{J}_x\rangle =  \sum_n a_n T^n |n\rangle,  \;\; |\mathsf{Q}_x\rangle =  \sum_n b_n T^n |n\rangle,\;\; \mathrm{etc.}   \label{eq:Tnn}
\end{equation}
Observing that
  \begin{equation}
\sum_{mn} a_m a_n T^{m+n} \langle m|\mathsf{W}^{-1}_{\mathrm{imp}}|n\rangle  \sim \sum_{mn} a_m a_n  T^{m+n + |m-n|} = \sum_{mn} a_m a_n  T^{2\max(m,n)},
\end{equation}
we conclude that to compute a transport coefficient up to $\mathrm{O}(T^{2k})$, we only need to keep basis vectors $|0\rangle,\ldots, |k\rangle$.     To leading order, we know from the Wiedemann-Franz law and Mott law that $T\bar\kappa$ and $T\alpha$ will both scale as $T^2$ at low temperatures.   As we  show in Section \ref{sec:strongint}, keeping only the $|0\rangle$ and $|1\rangle$ basis vectors is sufficient to recover these relations in the non-interacting limit.

Now that we understand the  form of the ``collisionless" electron-impurity scattering, let us discuss the form of the electronic collision integral.   Since we are interested in the hydrodynamic regime of  electron flow, we focus on theories  where momentum is conserved in all electron-electron collisions (there is no umklapp).  This means that the momentum vector  \begin{equation}
|\mathsf{P}_x\rangle =  p_{\mathrm{F}}|q^0\rangle + |q^1\rangle  \approx \sqrt{\frac{\nu}{d}}\left[p_{\mathrm{F}}|0\rangle + \frac{\mpi T}{\sqrt{3}}|1\rangle\right]  \label{eq:P01}
\end{equation}
is a null vector of $\mathsf{W}$.     One simple choice of $\mathsf{W}$ that is  consistent with this requirement is \begin{equation}
\mathsf{W} \approx \gamma \left(|1\rangle - \frac{\mpi T}{\sqrt{3}p_{\mathrm{F}}v_{\mathrm{F}}}|0\rangle\right)\left(\langle  1| - \frac{\mpi T}{\sqrt{3}p_{\mathrm{F}}v_{\mathrm{F}}}\langle 0|\right) + \sum_{n=2}^\infty \gamma |n\rangle\langle n|.  \label{eq:reltime}
\end{equation}
This choice, which is used extensively in transport theories, is often called the relaxation time approximation \cite{bgk}.   Note that the relative decay of longitudinal fluctuations above necessitates inelastic scattering processes.
The coefficient $\gamma \sim \tau_{\mathrm{ee}}^{-1}$  is associated with the rate at which electron-electron collisions relax \emph{longitudinal} fluctuations of the distribution function, proportional to $p_x$:  \begin{equation}
\gamma \equiv \tilde\alpha^2 \frac{T^2}{T_{\mathrm{F}}}.  \label{eq:T2def}
\end{equation} 
where $\tilde\alpha^2$, an effective coupling constant in the problem which we do not specify, is (roughly) proportional to the interaction constant.      In lower dimensions, it  is possible for this  scattering time to  be  enhanced  by factors of $\log (T_{\mathrm{F}}/T)$ \cite{wilkins, quinn, jungwirth, zheng96, zala, qiuzi}, and we will not worry about these logarithms in this paper.   

We note that the Fermi liquid prediction (\ref{eq:T2def}) for the inelastic scattering rate -- with suitable logarithms --  has been verified experimentally (see e.g. \cite{sqmurphy} for a direct verification in a 2D GaAs system).   The $T^2$ dependence in (\ref{eq:T2def}) is completely consistent with the classic Fermi liquid paradigm of well-defined quasiparticles.  Indeed,  our kinetic picture is perturbative in interactions and so a NFL does not emerge in our model in 2D or 3D electron liquids.  This statement is independent of $\tilde\alpha^2$, although for large enough $\tilde\alpha^2$ our kinetic theory is not valid.    Normal 3D metals are strongly interacting in the sense that $\tilde\alpha$ (also commonly called `$r_{\mathrm{s}}$' in interacting Coulomb systems) $\sim 6$.   However, because $T_{\mathrm{F}}  \sim 10^4$ K, $\gamma$ is still small relative to $k_{\mathrm{B}}T/\hbar$ (the scattering rate below which a quasiparticle picture does not make sense \cite{lucasrmp}).   In 2D GaAs hole systems, the  interaction coupling constant can be $\sim 40$ in dilute experimental systems, putting the system deep in the hydrodynamic regime \cite{noh}.

\subsection{The Strongly Interacting Limit}
\label{sec:strongint}
We are now finally ready  to study transport phenomena across the ballistic-to-hydrodynamic crossover.   To start off, let us work in a simple limit where 
\begin{equation}
\Gamma \ll \tilde\alpha^2 T_{\mathrm{F}}.  \label{eq:Kllgamma}
\end{equation}
We will relax this assumption in Section \ref{sec:weakint}.   This inequality can be satisfied in 2D systems experimentally, but not in the usual metals.   (\ref{eq:Kllgamma}) implies that $\gamma=\Gamma$ at temperatures $T\ll T_{\mathrm{F}}$, where an expansion in low order basis vectors ought to be sensible, even in the hydrodynamic regime.  Thus, we will work only to the lowest non-trivial order in $T/T_{\mathrm{F}}$ in $\mathsf{W}_{\mathrm{imp}}$, to capture the effects of interactions.    From the form of the Wiedemann-Franz law (\ref{eq:WF}), Mott relation (\ref{eq:mott}), and the definition  (\ref{eq:thermoelcondboltz}), we  conclude that it will suffice to keep track of only the vectors $|0\rangle$ and $|1\rangle$.    This is \emph{not} equivalent to  a  Taylor  series expansion in $T$, as $\gamma$ itself depends on $T$ via (\ref{eq:T2def}).   We will return to this point in Section \ref{sec:weakint}.

Converting the $|q^n\rangle$ basis to the $|n\rangle$ basis, the charge and heat current vectors are  \begin{subequations}\label{eq:JQ01}\begin{align}
|\mathsf{J}_x\rangle &\approx -e \sqrt{\frac{\nu}{d}} \left[ v_{\mathrm{F}} |0\rangle + \frac{a\mpi T}{\sqrt{3}v_{\mathrm{F}}}|1\rangle\right],  \\
|\mathsf{Q}_x\rangle &\approx \sqrt{\frac{\nu}{d}} \left[  \frac{\mpi^2 T^2}{3p_{\mathrm{F}}}|0\rangle + \frac{\mpi Tv_{\mathrm{F}}}{\sqrt{3}}|1\rangle\right].
\end{align}\end{subequations}
The net collision integral $\tilde{\mathsf{W}}$ is given by 
 \begin{equation}
\widetilde{\mathsf{W}} \approx \left(\frac{\mpi^2 T^2}{3p_{\mathrm{F}}^2v_{\mathrm{F}}^2} \gamma + \Gamma\right)  |0\rangle \langle 0|  +\frac{\mpi T}{\sqrt{3}v_{\mathrm{F}}}   \left(\Gamma_1 - \frac{\gamma}{p_{\mathrm{F}}}\right) ( |0\rangle\langle 1| + |1\rangle\langle 0| )   + (\Gamma+\gamma)|1\rangle\langle 1|
\end{equation}
Computing the thermoelectric conductivity matrices has reduced to simply inverting a $2\times 2$ matrix:
 \begin{equation}
\widetilde{\mathsf{W}}^{-1} \approx \frac{1}{\Gamma}  |0\rangle \langle 0|  - \frac{\mpi T}{\sqrt{3}v_{\mathrm{F}}}   \frac{\Gamma_1 - \frac{\gamma}{p_{\mathrm{F}}}}{\Gamma(\Gamma+\gamma)} ( |0\rangle\langle 1| + |1\rangle\langle 0| )   + \frac{\frac{\mpi^2 T^2}{3p_{\mathrm{F}}^2v_{\mathrm{F}}^2} \gamma + \Gamma}{\Gamma(\Gamma+\gamma)}|1\rangle\langle 1|
\end{equation}
In this step, we have made approximations consistent with the assumptions stated at the start of this  subsection.

First, let us study the limit $\gamma  = 0$, which corresponds to the $T\rightarrow 0$ limit.  After some algebra we obtain \begin{subequations}\begin{align}
\sigma &  \approx  \frac{\nu e^2 v_{\mathrm{F}}^2}{d\Gamma}, \\
\kappa\approx\bar\kappa &\approx  \frac{\mpi^2 T}{3} \frac{\nu v_{\mathrm{F}}^2}{d\Gamma} \equiv \kappa_{\mathrm{non-int}}, \\
\alpha &\approx -e\frac{\mpi^2 T\nu}{3d}  \left[-\frac{ v_{\mathrm{F}} \Gamma_1}{\Gamma^2} + \frac{  a }{\Gamma }+ \frac{  v_{\mathrm{F}} }{p_{\mathrm{F}}\Gamma}   \right]  \equiv \alpha_{\mathrm{non-int}}.
\end{align}\end{subequations}
In each case, we have only written down the leading order terms in the $T\rightarrow 0$ limit (where, by definition, we should recover the standard Fermi liquid WF law since interaction-induced scattering vanishes at the Fermi surface).   It is clear that we recover the Wiedemann-Franz law (\ref{eq:WF}).   The Mott relation (\ref{eq:mott}) is also obeyed, because 
\begin{align}
\frac{\partial \sigma}{\partial \mu} &= \frac{1}{dv_{\mathrm{F}}}\frac{\partial \sigma}{\partial p_{\mathrm{F}}} = \frac{\Omega_d}{(2\mpi\hbar)^d dv_{\mathrm{F}}} \frac{\partial}{\partial p_{\mathrm{F}}}\frac{p_{\mathrm{F}}v_{\mathrm{F}}}{\Gamma} =  \frac{\Omega_d}{(2\mpi\hbar)^d dv_{\mathrm{F}}} \left[\frac{v_{\mathrm{F}}}{\Gamma} - \frac{p_{\mathrm{F}}v_{\mathrm{F}}\Gamma_1}{\Gamma^2} + \frac{p_{\mathrm{F}}a}{\Gamma}\right] \notag  \\
&= -\frac{\nu v_{\mathrm{F}} \Gamma_1}{d\Gamma^2} + \nu\frac{  a }{d\Gamma }+ \nu\frac{  v_{\mathrm{F}} }{dp_{\mathrm{F}}\Gamma}.  \label{eq:mottcheck}
\end{align}

Having confirmed that our theory correctly reproduces the non-interacting ``collisionless" limit  of electron-impurity scattering, let  us crank up the electron-electron collision rate $\gamma$.   The first thing we observe is that  \begin{equation}
\sigma \approx \frac{\nu e^2 v_{\mathrm{F}}^2}{d\Gamma} + \mathrm{O}\left(T^2, T^2\gamma\right).
\end{equation}
This is an important result:  interactions affect  the electrical conductivity only weakly (i.e. at higher-order terms) through hydrodynamic corrections.  By contrast, as we show below, interaction does affect the thermal conductivity and the thermoelectric conductivity. 
The electrical conductivity is, in this approximation, essentially independent of electron-electron scattering.  

Next, let us discuss the thermal conductivity $\bar\kappa$.  In this case, we obtain \begin{equation}
\bar\kappa = \frac{\nu \mpi^2  v_{\mathrm{F}}^2 T}{3d  \Gamma} \left[ \frac{\frac{\mpi^2 T^2}{3p_{\mathrm{F}}^2v_{\mathrm{F}}^2} \gamma + \Gamma - 2 \frac{\mpi^2 T^2}{3p_{\mathrm{F}}v_{\mathrm{F}}^2} (\Gamma_1 - \frac{\gamma}{p_{\mathrm{F}}}) }{\Gamma+\gamma}  + \frac{\mpi^2 T^2}{3p_{\mathrm{F}}^2  v_{\mathrm{F}}^2} \right] \approx \kappa_{\mathrm{non-int}} \frac{\Gamma}{\Gamma+\gamma} + \bar\kappa_{\mathrm{int}} \frac{\gamma}{\Gamma+\gamma},
\end{equation}
with \begin{equation}
\bar\kappa_{\mathrm{int}} = \frac{4\mpi^4 T^3 \nu}{9dp_{\mathrm{F}}^2\Gamma}. \label{eq:kappaint}
\end{equation}
The thermoelectric conductivity is given by \begin{equation}
\alpha \approx -e\frac{\nu \mpi^2 T v_{\mathrm{F}}^2}{3d}  \frac{\frac{\Gamma+\gamma}{p_{\mathrm{F}}v_{\mathrm{F}}}  - \frac{1}{v_{\mathrm{F}}} (\Gamma_1 - \frac{\gamma}{p_{\mathrm{F}}}) + \frac{a}{v_{\mathrm{F}}^2} (\frac{\mpi^2 T^2}{3p_{\mathrm{F}}^2v_{\mathrm{F}}^2} \gamma + \Gamma) }{\Gamma(\Gamma+\gamma)} \approx \alpha_{\mathrm{non-int}} \frac{\Gamma}{\Gamma+\gamma} + \alpha_{\mathrm{int}} \frac{\gamma}{\Gamma+\gamma},   \label{eq:alpha33}
\end{equation}
where \begin{equation}
\alpha_{\mathrm{int}} = -e \frac{2\mpi^2 T \nu v_{\mathrm{F}}}{3d p_{\mathrm{F}} \Gamma}.  \label{eq:alphaint}
\end{equation}
Both $\bar\kappa$ and $\alpha$ admit an elegant interpretation in this limit:  they are a  weighted average of the ``collisionless" transport  coefficient and an ``interaction-limited" transport  coefficient.  The weights in  front of each term correspond to  the fraction of the  total scattering rate that is in the electron-impurity vs. electron-electron  channel.

We can go further.  In the interaction-dominated limit, we expect that the thermoelectric conductivity matrix is limited entirely by momentum relaxation \cite{lucasrmp, lucasreview17, hkms, hofman, lucasMM}.  Writing down the momentum conservation equation, we  obtain \begin{equation}
-en \mathbf{E} - s \nabla T = \Gamma_* \mathbf{v},
\end{equation}
with $\mathbf{v}$ the velocity of the interacting fluid, and $\Gamma_*$ a coefficient related to weak momentum relaxation.   The  left hand side corresponds to the momentum added to the system by the external drives;  the right hand side corresponds to the system's internal momentum dissipation.    Approximating the charge current as $\mathbf{J} \approx n\mathbf{v}$ and the heat current as $\mathbf{Q} \approx Ts\mathbf{v}$, we arrive at the formulas \begin{equation}
\left(\begin{array}{cc} \sigma &\ T\alpha \\ T \alpha &\ T\bar\kappa \end{array}\right) \approx \frac{1}{\Gamma_*} \left(\begin{array}{cc}  e^2n^2 &\ -eTsn \\ -eTsn &\ (Ts)^2 \end{array}\right)     \label{eq:hydrotrans}.
\end{equation}
Using the thermodynamic relations of a low temperature Fermi liquid given in  Appendix \ref{app:thermo},  we find that in our toy  model above, \begin{equation}
\Gamma_* = \mathcal{M} \Gamma,
\end{equation}
where $\mathcal{M}$ is the momentum-momentum susceptibility (analogous to a mass density).   Since $\Gamma$ is the momentum relaxation rate, this agrees with the predictions of the memory matrix formalism \cite{hofman, lucasMM}.
Using (\ref{eq:kappaint}) and (\ref{eq:alphaint}) with these thermodynamic identities, we can confirm that $\alpha_{\mathrm{int}}$  and $\bar\kappa_{\mathrm{int}}$ take the form demanded by (\ref{eq:hydrotrans}), relative to $\sigma$.   Note also the identity \begin{equation}
\sigma \bar\kappa_{\mathrm{int}} = T\alpha^2_{\mathrm{int}},   \label{eq:sigmakappaiden}
\end{equation}
which is tied to the fact that one process -- momentum relaxation -- limits all transport coefficients in the hydrodynamic regime.

Now, let us return to the fate of the Wiedemann-Franz law.   Conventionally one compares not $\bar\kappa$ to $\sigma$, but $\kappa$ to $\sigma$.  As a consequence, using the identity (\ref{eq:sigmakappaiden}), we obtain \begin{equation}
\kappa \approx \frac{\Gamma}{\Gamma+\gamma} \kappa_{\mathrm{non-int}}   + \frac{\gamma\Gamma}{(\Gamma+\gamma)^2} \left(\bar\kappa_{\mathrm{int}} - \frac{2T\alpha_{\mathrm{int}}\alpha_{\mathrm{non-int}}}{\sigma}\right) \approx \frac{\Gamma}{\Gamma+\gamma} \kappa_{\mathrm{non-int}} \approx \frac{\mpi^2 T \nu v_{\mathrm{F}}^2}{3d(\Gamma+\gamma)}  \label{eq:kappa}
\end{equation}  
As before, we are neglecting contributions that are subleading  in $T$, at each order in $\gamma$.   The Wiedemann-Franz law will be violated, ``strongly interacting" limit, in the particularly elegant manner given in (\ref{eq:33WF}).  
In this simple model, the amount of violation of the Wiedemann-Franz law corresponds directly to the fraction of scattering processes which are momentum-conserving.   This interpretation fails to hold if $T/T_{\mathrm{F}}$ is not so small, and/or if the disorder strength is comparable to the interaction strength (Section \ref{sec:weakint}).
Earlier discussions emphasizing that the Wiedemann-Franz law will be violated in the hydrodynamic limit of a finite density metal, because $\kappa/\sigma T$ is very small, can be found  in \cite{mahajan, vignale, muller2}.    In the hydrodynamic limit, the value of $\kappa$ matches the dissipative hydrodynamic coefficient $\kappa_{\textsc{q}}$, at leading order in $T/T_{\mathrm{F}}$: see Appendix \ref{app:kappa}.

The fate of the Mott relation is less dramatic.   In the limit $T\ll  T_{\mathrm{F}}$, both $\alpha_{\mathrm{non-int}}$ and $\alpha_{\mathrm{int}}$ are linear in $T$.   The collisionless-to-hydrodynamic crossover will simply correspond to a gradual change in the slope of $\alpha(T)$.    We note  that this crossover can be  more dramatic near charge neutrality ($T\gtrsim T_{\mathrm{F}}$).  Some experimental evidence for this effect in graphene, and further discussion, can be found in \cite{ghahari, foster2}.

\subsection{The Weakly Interacting Limit}
\label{sec:weakint}
In this section, we will consider the limit where (\ref{eq:Kllgamma}) is not obeyed.  This corresponds to metals with relatively low Fermi temperature (where the WF law is affected by thermal smearing of the Fermi function even without any interaction effects), and/or relatively small interaction coupling $\tilde\alpha$.  Indeed, smearing of the Fermi surface leads to comparable $\mathrm{O}\left(T^2\right)$ corrections, similar to (but distinct from) electron-electron interactions.   

To correctly compute the thermal conductivity, we must compute $T\bar\kappa$ to $\mathrm{O}(T^4)$, and so we must keep the basis vector $|2\rangle$ in our  expansion.  For simplicity, we will take the dispersion relation $\epsilon(p)$ to be exactly quadratic.  This means that in (\ref{eq:disp}), $b=0$, $ap_{\mathrm{F}} = v_{\mathrm{F}}$, and $p_{\mathrm{F}}v_{\mathrm{F}} = 2T_{\mathrm{F}}$.  We will also treat electron-electron collisions within a relaxation time approximation (\ref{eq:reltime}), but at next-to-leading order this does not change the form of the answer.  The forms of $|\mathsf{J}_x\rangle$, $|\mathsf{Q}_x\rangle$ and $\mathsf{W}_{\mathrm{imp}}$ are written explicitly in Appendix \ref{app:3basis}.    We then explicitly compute \begin{subequations}\label{eq:34eq}\begin{align}
\sigma &= \frac{\nu}{2} \frac{v_{\mathrm{F}}^2}{\Gamma} \left(1+ \frac{\mpi^2 T^2}{3 T_{\mathrm{F}}^2} \right) + \mathrm{O}\left(T^4\right), \\ 
\kappa &= \frac{\nu}{2} \frac{\mpi^2 T}{3e^2} \frac{v_{\mathrm{F}}^2}{\Gamma} \left(1 - \frac{17\mpi^2 T^2}{30T_{\mathrm{F}}^2} - \frac{\tilde\alpha^2 T^2}{\Gamma T_{\mathrm{F}}}\right)  + \mathrm{O}\left(T^5\right).
\end{align}\end{subequations}
The Lorenz number \begin{equation}
\frac{\kappa}{\sigma T}  = \frac{\mpi^2}{3e^2} \left[1 -\frac{9\mpi^2 T^2}{10T_{\mathrm{F}}^2} - \frac{\tilde\alpha^2 T^2}{\Gamma T_{\mathrm{F}}} \right]+\cdots
\end{equation}
is thus decreased both by thermal smearing of the Fermi surface and by electron-electron interactions.  When disorder is very weak,  the dominant effect will be that of electron-electron interactions.   Also observe that even the first subleading temperature-dependent correction to $\sigma$ arises only from thermal smearing  of the Fermi surface, and not from electron-electron scattering.

Thus, $\Gamma$, $\gamma$, and $T_{\mathrm{F}}$ are all relevant parameters determining the effective Lorenz number as a function of $T$.  Remembering that $T_{\mathrm{F}} \sim n^{2/d}$, where n is the electron density of the $d$-dimensional electron liquid,\footnote{This relation changes for systems with non-parabolic dispersion relations, such as monolayer graphene.} and that $\gamma \sim T^2/T_{\mathrm{F}}$, we conclude that lowering electron density may enhance both the effect of interaction and the effect of Fermi surface thermal smearing depending on the temperature of the system.  If the effective impurity disorder Gamma itself has a temperature dependence through electronic screening (as it does in 2D systems \cite{hwang15}), the situation gets much more complicated.   Disentangling all of these effects quantitatively will require materials-specific computations beyond the scope of the present paper.

\subsection{Hydrodynamic Fluctuations and Impurities}
\label{sec:eeimp}
In this section, we will qualitatively describe the  interplay of electron-electron interactions with the impurity potential.   Mathematically, this is done by keeping track of the full $\mathsf{W}$ matrix when performing the matrix inverse in (\ref{eq:Wtildedef}).    For the purposes of this section, we will keep the discussion qualitative, since the quantitative details depend on far too many unknown material-specific parameters and the precise nature of the disorder \cite{hartnoll1706, lucasRFB}.    After a microscopic calculation, what one finds is that $\mathsf{W}_{\mathrm{imp}}$ is no longer approximately a single scattering rate $\Gamma$, times the identity matrix.  Instead, the eigenvalues of $\mathsf{W}_{\mathrm{imp}}$ will sensitively depend on  the nature of disorder, and on the modes $|\alpha\rangle$ which are being sourced.   

We assume that the inhomogeneous potential $V_{\mathrm{imp}}(\mathbf{x})$ varies on the length scale $\xi$.   Let $\theta \ll 1$ denote the typical deflection angle of a quasiparticle after it moves a distance $\sim \xi$ through the potential landscape.     A typical degree of freedom $|\alpha\rangle$ will obey \begin{equation}
\Gamma \sim \langle \alpha | \mathsf{W}_{\mathrm{imp}} | \alpha\rangle  \sim \theta^2 \frac{v_{\mathrm{F}} \ell_{\mathrm{ee}}}{\xi (\xi + \ell_{\mathrm{ee}})},
\end{equation}
where \begin{equation}
\ell_{\mathrm{ee}} \equiv \frac{v_{\mathrm{F}}}{\gamma}
\end{equation}
This will hold, at least qualitatively, for all modes $|0\rangle$, $|1\rangle$, $|2\rangle$, etc., in the isotropic Fermi liquid with potential disorder \cite{hartnoll1706}.  In these theories, $\Gamma$  \emph{decreases}  as electron-electron interactions increase \cite{hartnoll1706}:  \begin{equation}
\Gamma \approx  \theta^2 \frac{v_{\mathrm{F}}}{\xi}  \left[ 1 - \frac{\xi \gamma}{v_{\mathrm{F}}}   + \cdots \right] \approx \theta^2 \frac{v_{\mathrm{F}}}{\xi}  - \theta^2 \gamma + \mathrm{O}\left(\gamma^2\right),  \label{eq:35Gamma1}
\end{equation}
   We conclude from (\ref{eq:33WF}) that the Wiedemann-Franz law is violated even more strongly when hydrodynamic corrections to impurity scattering are accounted for, since the effective disorder strength $\Gamma$ is suppressed by interactions.    From (\ref{eq:35Gamma1}), we also estimate that the hydrodynamic corrections to  impurity scattering are relatively weak compared to intrinsic electron-electron scattering in the collision integral by a factor $\theta^2$.  This arises from the inherent difference between elastic (impurity) and inelastic (electron-electron) scattering processes.
   
   However, in some cases it is possible for hydrodynamic corrections to impurity scattering to  behave rather differently.  In particular, in the presence of ``magnetic" disorder,  one obtains \cite{lucasRFB} \begin{equation}
   \langle 0 | \mathsf{W}_{\mathrm{imp}} | 0\rangle \approx \theta^2 v_{\mathrm{F}} \left(\frac{1}{\xi} + \frac{1}{\ell_{\mathrm{ee}}}\right)  =  \theta^2 \frac{v_{\mathrm{F}}}{\xi} + \theta^2\gamma  \equiv \Gamma_*.
   \end{equation}
  $|1\rangle$, $|2\rangle$, etc. will have the same decay rate as before.    The origin of the peculiar behavior of the decay rate of the $|0\rangle$ mode is related to the fact that the ``magnetic" disorder exerts large shear stresses  which in turn create large local velocity fields in the electronic fluid.   We must now recompute $\sigma$ and $\kappa$ in a fluid where \begin{equation}
  \mathsf{W}_{\mathrm{imp}} =  \Gamma_*  |0\rangle\langle  0| + \Gamma|1\rangle\langle 1|.
  \end{equation}
  Following the procedure of Section \ref{sec:strongint}, we obtain \begin{subequations}\begin{align}
  \sigma &\approx \frac{\nu}{d}  \frac{e^2 v_{\mathrm{F}}^2}{\Gamma_*}, \\
  \kappa &\approx  \frac{\nu}{d}  \frac{\mpi^2 T}{3} \frac{v_{\mathrm{F}}^2}{\Gamma+\gamma}.
  \end{align}\end{subequations}
  We therefore find that for this fluid, \begin{equation}
  \theta^2 \lesssim \frac{3e^2}{\mpi^2} \frac{\kappa}{\sigma T} \le 1.
  \end{equation}
  The Wiedemann-Franz law is still violated, but not as strongly as before.  Furthermore, the Lorenz ratio will no longer become arbitrarily small as the sample becomes more pure.    This scenario is analogous to the  presence of electron-electron umklapp (momentum-relaxing) scattering at rate $\theta^2\gamma$.

Finally, another well-known \cite{hwang15} example of an interaction-suppressing effect of disorder is the screening of long range Coulomb impurities by the electrons themselves, which leads to a strongly temperature-dependent effective disorder in 2D systems, manifesting a first-order positive thermal correction to the resistivity which is linear in $T/ T_{\mathrm{F}}$ in 2D (but not in 3D).  This effect, which is beyond the scope of the current hydrodynamic theory as it arises from the singular nature of the 2D polarizability function at $2k_{\mathrm{F}}$ \cite{stern, dassarma86},  tends to increase $\Gamma$, and therefore the Lorenz number, with increasing temperature.   As this effect will increase the Lorenz number it does not provide an alternate explanation for reduced Lorenz numbers in an interaction-limited regime.

\section{Magnetotransport}
\label{sec:magnetic}
In this section, we will discuss magnetotransport phenomena in the presence of a weak magnetic field, focusing  specifically on the experimentally relevant case of a two-dimensional electron  fluid.   When the cyclotron radius is very large compared to $\lambda_{\mathrm{F}}$, the magnetic field can  be treated within our classical kinetic description.   If the cyclotron radius is large compared to the size of the impurity potential, then we can neglect magnetic field corrections to $\mathsf{W}_{\mathrm{imp}}$.   We will make both of these assumptions to simplify the calculations below.

Accounting for the magnetic field amounts to modifying $\widetilde{\mathsf{W}}$ to $ \widetilde{\mathsf{W}} = \mathsf{W}+\mathsf{W}_{\mathrm{imp}}+ \mathsf{W}_{\mathrm{mag}}$, with \begin{equation}
\mathsf{W}_{\mathrm{mag}} = -eB \epsilon_{ij} v_i \frac{\partial}{\partial p_j}
\end{equation}
and $\epsilon_{xy}=-\epsilon_{yx}=1$ the antisymmetric Levi-Civita tensor.    In a rotationally invariant Fermi liquid, we can further simplify this.   Generalizing (\ref{eq:alphabasis}) to \begin{equation}
|\alpha_i\rangle \equiv \frac{p_i}{p} f_\alpha(p),
\end{equation}
and generalizing the bases discussed in Section \ref{sec:basis}, we obtain that \begin{equation}
\mathsf{W}_{\mathrm{mag}} | \alpha_i \rangle = e\epsilon_{ij}\frac{p_j}{p}  \frac{Bv(p)}{p} f_\alpha (p);
\end{equation}
hence, $\mathsf{W}_{\mathrm{mag}}$ will mix $x$ and $y$ vectors (under spatial rotation), but will not couple vectors to higher/lower rank tensors.

\subsection{The Strongly Interacting Limit}
For simplicity, let us now focus on  the strongly interacting limit described in Section \ref{sec:strongint}.   As before, we will (up to the $T$-dependence of $\gamma$) only be interested in the leading  order temperature dependence of all conductivities, which will allow us to neglect all basis vectors but $|0_{x,y}\rangle$ and $|1_{x,y}\rangle$.   We may  write \begin{equation}
\widetilde{\mathsf{W}} = \left(\mathsf{W}+\mathsf{W}_{\mathrm{imp}}\right) \otimes \mdelta_{ij} + \widetilde{\mathsf{W}}_{\mathrm{mag}}  \otimes  \epsilon_{ij},
\end{equation}
where \begin{equation}
\widetilde{\mathsf{W}}_{\mathrm{mag}} =  \frac{eBv_{\mathrm{F}}}{p_{\mathrm{F}}} \left(|0\rangle \langle 0| + |1\rangle\langle 1|\right) + \frac{\mpi T eB(ap_{\mathrm{F}} - v_{\mathrm{F}})}{\sqrt{3}v_{\mathrm{F}}p_{\mathrm{F}}^2} (|0\rangle\langle 1| + |1\rangle\langle 0|).
\end{equation}
Using (\ref{eq:thermoelcondboltz}), we can invert the $4\times 4$ matrix $\widetilde{\mathsf{W}}$ to compute all conductivities of interest.    Taking the same limit as Section  \ref{sec:strongint}, we find that the electrical conductivity is given by 
\begin{subequations}\begin{align}
\sigma_{xx} = \sigma_{yy} &= \frac{\nu}{d} \frac{e^2 v_{\mathrm{F}}^2\Gamma}{\Gamma^2+\omega_{\mathrm{c}}^2}, \\
\sigma_{xy} &= -\sigma_{yx} = \frac{\nu}{d} \frac{e^2 v_{\mathrm{F}}^2\omega_{\mathrm{c}}}{\Gamma^2+\omega_{\mathrm{c}}^2},
\end{align}\end{subequations}
where we have defined the cyclotron frequency \begin{equation}
\omega_{\mathrm{c}} \equiv \frac{eBv_{\mathrm{F}}}{p_{\mathrm{F}}}.
\end{equation}
As before, we observe that $\sigma_{ij}$ is  independent of electron-electron interactions, within this simple model.   Furthermore, the form of $\sigma_{ij}$ is completely consistent with the Drude model of magnetotransport: see e.g. \cite{lucasreview17}.   The resistivity tensor $\rho_{ij} =  \sigma^{-1}_{ij}$ exhibits no classical magnetoresistance, following the conventional lore.   When $\Gamma=0$, we obtain the Hall conductivity \begin{equation}
\sigma_{xy} = \frac{\nu}{2} \frac{e^2 v_{\mathrm{F}}^2}{\omega_{\mathrm{c}}}  = \frac{en}{B}. 
\end{equation}
Since $-en$ is the  charge density, we find the classical Hall conductivity.  In fact, using quantum Ward identities, this relation can be derived for any translation invariant quantum system  (in the absence of Berry curvature):  see e.g. \cite{lucasrmp}.
The open-circuit thermal conductivities are given by \begin{subequations}\label{eq:magWF}\begin{align}
\kappa_{xx} = \kappa_{yy} &= \frac{\nu}{d} \frac{\mpi^2 T}{3} \frac{v_{\mathrm{F}}^2(\Gamma+\gamma)}{(\Gamma+\gamma)^2+\omega_{\mathrm{c}}^2}, \\
\kappa_{xy} = -\kappa_{yx} &= -\frac{\nu}{d} \frac{\mpi^2 T}{3} \frac{v_{\mathrm{F}}^2 \omega_{\mathrm{c}}}{(\Gamma+\gamma)^2+\omega_{\mathrm{c}}^2},
\end{align}\end{subequations}
while the closed-circuit thermal conductivity is \begin{subequations}\begin{align}
\bar\kappa_{xx} = \bar\kappa_{yy} &\approx \kappa_{xx} + \frac{\nu}{d} \frac{4\mpi^4  T^3  \Gamma \gamma^2}{p_{\mathrm{F}}^2(\Gamma^2 + \omega_{\mathrm{c}}^2)((\Gamma+\gamma)^2+\omega_{\mathrm{c}}^2)}, \\
\bar\kappa_{xy} = -\bar\kappa_{yx} &\approx \kappa_{xy} -  \frac{\nu}{d} \frac{4\mpi^4  T^3  \omega_{\mathrm{c}} \gamma^2}{p_{\mathrm{F}}^2(\Gamma^2 + \omega_{\mathrm{c}}^2)((\Gamma+\gamma)^2 + \omega_{\mathrm{c}}^2)}.
\end{align}\end{subequations}
The thermoelectric conductivity is given by \begin{subequations}\label{eq:magmott}\begin{align}
\alpha_{xx} = \alpha_{yy} &= - \frac{e\nu}{d} \frac{\mpi^2 T}{3}    \frac{\Gamma (\Gamma+\gamma)(ap_{\mathrm{F}}\Gamma + v_{\mathrm{F}}(\Gamma -  \Gamma_1 p_{\mathrm{F}} + 2\gamma)) + \omega_{\mathrm{c}}^2 (3\Gamma v_{\mathrm{F}} + \Gamma_1 p_{\mathrm{F}} v_{\mathrm{F}}-a\Gamma p_{\mathrm{F}})}{p_{\mathrm{F}}(\Gamma^2 +  \omega_{\mathrm{c}}^2)((\Gamma+\gamma)^2 +  \omega_{\mathrm{c}}^2)}, \\
\alpha_{xy} = \alpha_{yx} &= \frac{e\nu}{d} \frac{\mpi^2 T}{3}    \frac{p_{\mathrm{F}}(a\Gamma-\Gamma_1 v_{\mathrm{F}})(2\Gamma+\gamma) + v_{\mathrm{F}}(2\gamma^2 + 2\omega_{\mathrm{c}}^2 + 3\Gamma\gamma)}{p_{\mathrm{F}}(\Gamma^2 +  \omega_{\mathrm{c}}^2)((\Gamma+\gamma)^2 +  \omega_{\mathrm{c}}^2)}.
\end{align}\end{subequations}

We now unpack the thermal and thermoelectric conductivities.   We first focus on the non-interacting limit $\gamma=0$.  
 The Wiedemann-Franz law is obeyed component-wise:  \begin{equation}
\bar\kappa_{ij} \approx \kappa_{ij} \approx  \frac{\mpi^2 T}{3 e^2} \sigma_{ij},
\end{equation}
as is the Mott relation:  \begin{equation}
\alpha_{ij} = -\frac{\mpi^2 T}{3 e}\frac{\partial \sigma_{ij}}{\partial \mu}.
\end{equation}
This relation can be checked using methods analogous  to (\ref{eq:mottcheck}).   In the limit of large $\gamma$, we find that \begin{subequations}\label{eq:sigmahydromagnetic}\begin{align}
\left(\begin{array}{cc} \sigma_{xx} &\  T\alpha_{xx} \\ T\alpha_{xx} &\ T\bar\kappa_{xx} \end{array}\right) &= \frac{1}{\mathcal{M}} 
\left(\begin{array}{cc}  e^2n^2 &\ -eTsn \\ -eTsn &\ (Ts)^2 \end{array}\right)     \frac{\Gamma}{\Gamma^2+\omega_{\mathrm{c}}^2} = \left(\begin{array}{cc}  e^2n^2 &\ -eTsn \\ -eTsn &\ (Ts)^2 \end{array}\right)     \frac{\mathcal{M}\Gamma}{(\mathcal{M}\Gamma)^2+(enB)^2} \\
\left(\begin{array}{cc} \sigma_{xy} &\  T\alpha_{xy} \\ T\alpha_{xy} &\ T\bar\kappa_{xy} \end{array}\right) &= \frac{1}{\mathcal{M}} 
\left(\begin{array}{cc}  e^2n^2 &\ -eTsn \\ -eTsn &\ (Ts)^2 \end{array}\right)     \frac{-\omega_{\mathrm{c}}}{\Gamma^2+\omega_{\mathrm{c}}^2}= \left(\begin{array}{cc}  e^2n^2 &\ -eTsn \\ -eTsn &\ (Ts)^2 \end{array}\right)     \frac{-enB}{(\mathcal{M}\Gamma)^2+(enB)^2}.
\end{align}\end{subequations}
These latter equations are consistent with the predictions of the memory matrix formalism at weak disorder and weak magnetic field strength \cite{lucasMM}.   

When $\Gamma$, $\gamma$ and $\omega_{\mathrm{c}}$ are all comparable, we conclude from (\ref{eq:magWF}) and (\ref{eq:magmott}) that the Wiedemann-Franz and Mott relations are no longer valid.   The thermal conductivity depends rather simply on interactions: to leading order in $T/T_{\mathrm{F}}$, we may simply replace $\Gamma \rightarrow \Gamma+\gamma$, as in Section \ref{sec:strongint}.  However, the thermoelectric conductivity $\alpha_{ij}$ is far more complicated, and we do not see a simple way to disentangle the effects of finite magnetic field, disorder and interaction strength.

\section{Discussion and Conclusions}
\label{sec:conclusion}
The WF and Mott laws are strongly affected by electron-electron interactions in the hydrodynamic regime entirely within the Fermi liquid paradigm.  Even an arbitrary suppression of the WF ratio well below the ideal Lorenz number does not necessarily signify any NFL, but may only indicate strong interaction effects in the hydrodynamic regime \cite{vignale}.   The fact that most normal 3D metals manifest the ideal WF behavior is a consequence of the fact that  $\Gamma\gg\gamma$  by virtue of the large $T_{\mathrm{F}} \sim 10^{4}$--$10^5$ K, and also the relative strength of electron-phonon scattering.   
This is in spite of 3D normal metals being 'strongly interacting' in the sense of having a large dimensionless coupling constant $r_{\mathrm{s}}\sim 6$.   At low temperatures, where electron-phonon interaction is negligible compared with electron-electron interaction, the electron-impurity scattering is stronger than electron-electron interaction.   We  do not see any obvious theoretical obstruction to a 3D metal where electronic hydrodynamics occurs so long as the condition $\gamma  > \Gamma$ exists;  a recent experiment suggests such 3D systems exist \cite{felser}.


We expect that at least some of the experimentally observed strange metals are in a hydrodynamic regime.  Earlier discussion of this point may also be found in \cite{dsz, hartnoll1704}.   Is it also possible that this hydrodynamic regime is reasonably described by the Fermi liquid approach detailed in this paper?
Answering this question requires careful material-specific considerations of each system to investigate (\emph{i}) whether $\gamma\gg\Gamma$ is satisfied, and (\emph{ii}) if the failure of the WF law is happening in a temperature range consistent with our Eq. (\ref{eq:33WF}).  Such an investigation is well beyond the scope of the current work, but is worth future consideration.  Perhaps (at least in some portions of an often complicated phase diagram) strange metals are 'strange' only in the sense of being hydrodynamic metals.   Indeed, most strange metals have rather low $T_{\mathrm{F}}\sim 10^3$ K, so strong interaction effects could easily drive the system into the hydrodynamic regime.    

In many strange metals, the large $T$-linear resistivity, violating the Mott-Ioffe-Regel `bound' \cite{takagi}, is used as additional evidence for NFL physics \cite{mackenzie2013, hartnoll1}.   The validity of our kinetic approach is suspect in this limit.   Nevertheless, building upon \cite{hartnoll1704}, we propose a more careful study of thermal and electrical transport in the Fermi liquid regime of strange metals, where ``conventional" $T^2$ resistivity is observed just outside of the ``critical fan" of $T$-linear resistivity, and where the kinetic theory of transport should apply.   It is reasonable to assume that disorder has the same origin in both strange and conventional portions of the phase diagram \cite{hartnoll1704}, so we expect that a clear understanding of the roles  of umklapp, phonon scattering, and hydrodynamic effects in this regime will shed light on the origin of $T$-linear  resistivity.

All solid state materials have (at least) three distinct scattering mechanisms affecting transport:  electron-impurity scattering (controlling low-temperature transport), electron-phonon scattering (controlling high-temperature transport), and electron-electron scattering (controlling the ballistic to hydrodynamic crossover, in the absence of umklapp).   Hydrodynamics can only be observed if $\gamma \sim T^2/T_{\mathrm{F}}$ is the dominant of these three scattering mechanisms.\footnote{We cannot formally rule out momentum-conserving electron-phonon scattering as playing an important role, but this seems unlikely in most materials.}  This rules out hydrodynamic observations at high temperatures, since electron-phonon scattering rates (not considered in our work) scaling as $\sim T/T_{\mathrm{D}}$ (for $T>T_{\mathrm{D}}$), where $T_{\mathrm{D}}$ ($\sim10^2$ -- $10^3$ K) is the typical phonon energy scale in most materials, is the dominant resistive scattering mechanism in all electronic materials at higher temperatures.  At low temperatures, the requirement $\gamma \gg \Gamma$ necessitates both a very pure system so that the impurity disorder strength is weak and also a Fermi temperature low enough for $\gamma$ not to be too small.  For a Coulomb-interacting Fermi liquid with parabolic band dispersion with an effective mass $m$,  the requirement for hydrodynamics becomes  $m/n^{2/d}$ being very large, where $n$ is the carrier density and $d$ the spatial dimension.  In addition, the "low-temperature" condition $T\ll T_{\mathrm{D}}$ must be satisfied in order for phonon scattering to be unimportant.  It turns out, as emphasized already, this quantity is rather small in 3D metals by virtue of $n$ being too large.  However, in 2D GaAs systems, the hydrodynamic condition can easily be satisfied, both because the carrier density can be made very low and the impurity disorder can be made very small by utilizing modulation doping.  

Indeed, the systems where we believe that our hydrodynamic predictions are most likely going to be verified are low-disorder, high-mobility 2D systems such as modulation-doped 2D GaAs systems and high-quality graphene layers.  Converting the available experimental 2D mobility values and theoretical (but experimentally verified) electron-electron scattering strengths to effective scattering times, we find that the condition $\gamma \gg \Gamma$ is obeyed in 2D GaAs structures (both n-doped and p-doped) down to 1 K or below in dilute systems of density $\sim10^{11} \; \mathrm{cm}^{-2}$  \cite{lilly, manfra}.    At these low temperatures, phonon effects are completely negligible, and therefore, our approximations should apply uncritically with the dilute 2D GaAs system being deep in the hydrodynamic regime.  We predict a strong failure of the WF law in these systems, and our prediction of (\ref{eq:33WF}) can be directly verified by varying the temperature.   To be specific, in the 2D n-GaAs system (assuming a mobility of $10^7\; \mathrm{cm}^2/\mathrm{V}\cdot\mathrm{s}$), the electron-electron and the electron-impurity scattering times at $T=3$ K and $n=3\times 10^{11} \mathrm{cm}^{-2}$  are respectively $\tau_{\mathrm{ee}}=1/\gamma=20$ ps and $\tau_{\mathrm{ei}} =1/\Gamma=400$ ps, making $\gamma \sim 20\Gamma$, thus placing the system deep into the hydrodynamic regime.  Lowering $T$ to 100 mK in the same sample hardly modifies the mobility (i.e. $\Gamma$), but $\tau_{\mathrm{ee}}$ increases to 25000 ps, driving the system  to $\gamma \ll  \Gamma$.  Since $T_{\mathrm{F}} = 125 \; \mathrm{K}$ for $n=3\times 10^{11} \mathrm{cm}^{-2}$, the low-temperature condition necessary for the applicability of our Eq. (\ref{eq:33WF}) remains valid throughout.  Thus, changing temperature from 3 K to 100 mK in a high mobility 2D n-GaAs system would be an ideal experiment for the verification of our theory of the failure of the WF law.  The same is even more true for 2D p-GaAs holes also, since the hole mass is larger than the electron mass in GaAs, thus making $\gamma$ effectively larger at the same temperature since  $T_{\mathrm{F}}~ 1/m$.  In both cases, electron-phonon interaction is at least an order of magnitude weaker than even the electron-impurity interaction at these cryogenic temperatures, making the high-mobility  2D GaAs structures the ideal systems for studying hydrodynamic transport effects.

Unfortunately, we find that the other well-known 2D semiconductor system, namely Si MOSFETs \cite{hu15}, does not satisfy the hydrodynamic condition (it comes close, but falls below the minimal requirement of $\gamma  \gtrsim \Gamma$) in any accessible density or temperature range.   Even the purest 2D Si system is simply not pure enough, at present.

In graphene, one has to go rather close to the Dirac point ($n\sim 10^9 \; \mathrm{cm}^{-2}$) to achieve the hydrodynamic constraint at $T\sim50$ K.  Here phonon effects are weak, but not negligible.  One therefore requires extremely clean samples so that impurity-induced puddle effects do not overwhelm the hydrodynamic effects at such low densities.  Such experiments have been performed and interpreted based on hydrodynamic theories recently \cite{lucas3, crossno}, but more work should be done to directly verify our predictions.    Recently, the relevance of clean bilayer graphene in the context of 2D hydrodynamic transport has been pointed out \cite{adam18}. 

In this paper, we have focused on the limit $T\ll T_{\mathrm{F}}$.  Assuming reasonable forms for the electron-electron interactions, in this limit we rigorously reduced the calculation of thermoelectric conductivities to a finite-dimensional linear algebra problem.   We emphasize,  however, that the techniques described in this paper immediately and straightforwardly generalize to $T\sim T_{\mathrm{F}}$ (so long as long-lived quasiparticles exist), and to systems without rotational invariance.   In these more complicated settings, it is likely that the evaluation of the thermoelectric conductivity matrix must be done numerically, following the theoretical formalism described in this work.

\addcontentsline{toc}{section}{Acknowledgements}
\section*{Acknowledgements}
AL was supported by the Gordon and Betty Moore Foundation's EPiQS Initiative through Grant GBMF4302.   SDS was supported  by Laboratory for Physical Sciences.

\begin{appendix}
\section{Low Temperature Expansion}
\label{app:details}
In this appendix, we describe more explicit details about  the low temperature expansion of inner products in the $|q^n\rangle$ and $|n\rangle$ bases, in Section \ref{sec:largeFS}.   We employ the integral \begin{equation}
\int\limits_{-\infty}^\infty \frac{\mathrm{d}x}{(1+\mathrm{e}^x)^n}  \mathrm{e}^{(m+1)x} = \frac{\mathrm{\Gamma}(1+m)\mathrm{\Gamma}(n-m-1)}{\mathrm{\Gamma}(n)} \equiv \mathcal{F}(m,n),
\end{equation}
Ignoring the singularity at the origin of momentum space only leads to exponentially small corrections ($\mathrm{e}^{-T_{\mathrm{F}}/T}$) in our theory.   We then define \begin{equation}
G(m,n,p) \equiv \frac{\partial^p \mathcal{F}(m,n)}{\partial m^p} = \int\limits_{-\infty}^\infty \frac{\mathrm{d}x}{(1+\mathrm{e}^x)^n}  \mathrm{e}^{(m+1)x}  x^p.
\end{equation}
Having defined $G(m,n,p)$, all integrals over momentum space in the inner products can be analytically evaluated.   But, the process is a little  bit subtle.   In particular, observe that the derivative  of the Fermi function must be carefully expanded:  \begin{equation}
\frac{ \mathrm{e}^{\epsilon(q)/T}}{T(1+\mathrm{e}^{\epsilon(q)/T})^2} = \frac{\mathrm{e}^{v_{\mathrm{F}}q/T}}{(1+\mathrm{e}^{v_{\mathrm{F}}q/T})^2}\left[1-\frac{(\mathrm{e}^{v_{\mathrm{F}}q/T}-1)}{(1+\mathrm{e}^{v_{\mathrm{F}}q/T})}\left(\frac{ap^2}{2}+\frac{bp^3}{3}\right) + -\frac{(\mathrm{e}^{2v_{\mathrm{F}}q/T}-4\mathrm{e}^{v_{\mathrm{F}}q/T}+1)}{2(1+\mathrm{e}^{v_{\mathrm{F}}q/T})^2}\left(\frac{ap^2}{2}+\frac{bp^3}{3}\right)^2 \right] +  \cdots
\end{equation}
This means that
\begin{align}
\langle q^n | q^m\rangle \approx \frac{\nu}{d}  \int\limits_{-\infty}^\infty \mathrm{d}q \left(1+\frac{q}{p_{\mathrm{F}}}\right)^{d-1} \frac{ \mathrm{e}^{\epsilon(q)/T}}{T(1+\mathrm{e}^{\epsilon(q)/T})^2}  q^{m+n}.
\end{align}
is not straightforwardly  given by something proportional to $G(0,2,m+n)$.  
In $d=2$, for example, we find the complicated Taylor series \begin{align}
\langle q^n | q^m\rangle &\approx \frac{\nu}{d} \left(\frac{T}{v_{\mathrm{F}}}\right)^{m+n} \left[G(0,2,m+n) - \frac{aT}{2v_{\mathrm{F}}^2} (G(1,3,m+n+2)-G(0,3,m+n+2)) \right. \notag \\
&\left.- \frac{bT^2}{3v_{\mathrm{F}}^3} (G(1,3,m+n+3)-G(0,3,m+n+3)) \right. \notag  \\
&\left. + \frac{a^2T^2}{8v_{\mathrm{F}}^4}(G(2,4,m+n+4)-4G(1,4,m+n+4)+G(0,4,m+n+4)) \right. \notag \\
&\left. + \frac{T}{v_{\mathrm{F}}p_{\mathrm{F}}}  \left(G(0,2,m+n+1) - \frac{aT}{2v_{\mathrm{F}}^2} (G(1,3,m+n+3)-G(0,3,m+n+3))\right)\right] +  \mathrm{O}\left(T^3\right).
\end{align}
The expansion of the inner product to $\mathrm{O}(T^2)$, as  above, will be adequate for the purposes of this paper.

\section{Thermodynamic Identities}
\label{app:thermo}
In this appendix, we describe the relationship between entropy, charge and the inner products in our  kientic theory.

First we re-derive an identity observed in \cite{hartnoll1706}: \begin{equation}
\langle \mathsf{J}_x | \mathsf{P}_x\rangle = \int \frac{\mathrm{d}^dp}{(2\mpi\hbar)^d} \left(-\frac{\partial f_{\mathrm{eq}}}{\partial \epsilon}\right) v_x p_x = \int \frac{\mathrm{d}^dp}{(2\mpi\hbar)^d} \left(-\frac{\partial f_{\mathrm{eq}}}{\partial p_x}\right) p_x = \int \frac{\mathrm{d}^dp}{(2\mpi\hbar)^d}  f_{\mathrm{eq}} = n.   \label{eq:JPinnerproduct}
\end{equation}
Observe  that the zero temperature  charge density is given by \begin{equation}
n_0 = \frac{p_{\mathrm{F}}^d}{d(2\mpi\hbar)^d}\Omega_d,
\end{equation}
where $\Omega_d$ denotes the volume of the unit sphere in $d$ spatial dimensions.  
Secondly, the density of states is given by \begin{equation}
\nu = \frac{\partial n_0}{\partial \mu} = \frac{1}{v_{\mathrm{F}}} \frac{\partial n_0}{\partial  p_{\mathrm{F}}} = \Omega_d \frac{p_{\mathrm{F}}^{d-1}}{v_{\mathrm{F}} (2\mpi\hbar)^d},
\end{equation}
In the large Fermi surface limit, in $d=2$, we have  \begin{equation}
n \approx \frac{\nu v_{\mathrm{F}}p_{\mathrm{F}}}{2}.
\end{equation}

Next, we find \begin{equation}
\langle \mathsf{Q}_x|\mathsf{P}_x\rangle = \int \frac{\mathrm{d}^dp}{(2\mpi\hbar)^d} \left(-\frac{\partial f_{\mathrm{eq}}}{\partial \epsilon}\right) v_x \epsilon p_x = \int \frac{\mathrm{d}^dp}{(2\mpi\hbar)^d} \left(-\frac{\partial f_{\mathrm{eq}}}{\partial p_x}\right) p_x \epsilon = \int  \frac{\mathrm{d}^dp}{(2\mpi\hbar)^d}  f_{\mathrm{eq}} \left(\epsilon + p_x v_x\right).
\end{equation}
The pressure is given by \begin{equation}
P =  \int  \frac{\mathrm{d}^dp}{(2\mpi\hbar)^d}  T \log \left(1+\mathrm{e}^{-(\epsilon-\mu)/T}\right)
\end{equation} and  we find (using the same integration by parts tricks as before) \begin{equation}
\int  \frac{\mathrm{d}^dp}{(2\mpi\hbar)^d}  f_{\mathrm{eq}}p_x v_x = \int  \frac{\mathrm{d}^dp}{(2\mpi\hbar)^d} T \log \left(1+\mathrm{e}^{-\epsilon/T}\right).
\end{equation}
We conclude that \begin{equation}
\langle \mathsf{Q}_x|\mathsf{P}_x\rangle  =  \langle  \epsilon\rangle + P =  Ts,  \label{eq:QPinnerproduct}
\end{equation}
where $s$ is the entropy density.  In the last step, we have used the Gibbs-Duhem identity together with the fact that we have fixed $\mu=0$ by our conventions.   In a Fermi liquid, we also have \begin{equation}
s  = \left(\frac{\partial P}{\partial T}\right)_\mu \approx  \frac{\mpi^2T}{3} \nu.
\end{equation}

Finally, the momentum-momentum susceptibility is given by \begin{equation}
\mathcal{M} = \lim_{v_x \rightarrow 0} \frac{1}{v_x} \int \frac{\mathrm{d}^d\mathbf{p}}{(2\mpi\hbar)^d} p_x f_{\mathrm{eq}}(\epsilon  - p_x v_x) =  \int \frac{\mathrm{d}^d\mathbf{p}}{(2\mpi\hbar)^d} \frac{p^2}{d} \left(-\frac{\partial f_{\mathrm{eq}}}{\partial \epsilon} \right) \approx \frac{\nu p_{\mathrm{F}}^2}{d}.
\end{equation}
Note that (in $d=2$, at least) \begin{equation}
\frac{n}{\mathcal{M}} = \frac{v_{\mathrm{F}}}{p_{\mathrm{F}}}.
\end{equation}
This identity allows one to show (\ref{eq:sigmahydromagnetic}).

\section{The First  Three Basis Vectors}
\label{app:3basis}
We must keep track of $|0\rangle$, $|1\rangle$ and $|2\rangle$, as well as keep track of $\mathrm{O}\left(T^2\right)$ corrections to $\mathsf{W}_{\mathrm{imp}}$, in order to correctly compute (\ref{eq:34eq}).  The vectors \begin{subequations}\begin{align}
|\mathsf{J}_x\rangle &\approx  \left(v_{\mathrm{F}} - \frac{\mpi^2 T^2}{6p_{\mathrm{F}}^2v_{\mathrm{F}}^3}\right)|0\rangle + \frac{\mpi T}{\sqrt{3}p_{\mathrm{F}}}|1\rangle, \\
|\mathsf{Q}_x\rangle &\approx \left(\frac{\mpi^2 T^2}{3p_{\mathrm{F}}} + \frac{21\mpi^4 T^4}{8p_{\mathrm{F}}^3 v_{\mathrm{F}}^2}\right)|0\rangle + \left(\frac{\mpi Tv_{\mathrm{F}}}{\sqrt{3}} - \frac{31\mpi^3 T^3}{15\sqrt{3}p_{\mathrm{F}}^2 v_{\mathrm{F}}}\right)|1\rangle + \left(\frac{2\mpi^2 T^2}{\sqrt{5}p_{\mathrm{F}}} + \frac{149 \mpi^4 T^4}{15\sqrt{5}p_{\mathrm{F}}^3 v_{\mathrm{F}}^2}\right)|2\rangle. \label{eq:QxC}
\end{align}\end{subequations}
We have included the first non-trivial subleading corrections in $T$, which will be necessary to compute the first non-trivial subleading corrections to conductivity.   As noted in Section \ref{sec:weakint}, we have assumed a quadratic dispersion relation; hence $2T_{\mathrm{F}} = p_{\mathrm{F}}v_{\mathrm{F}}$,  and $ap_{\mathrm{F}} = v_{\mathrm{F}}$.  Using (\ref{eq:Gammadef}) and the fact from (\ref{eq:alphaWimpbeta}) that $\mathcal{A}_{\mathrm{imp}}$ is a constant, we obtain  \begin{subequations}\begin{align}
\frac{\Gamma_1}{\Gamma} &= -\frac{2}{p_{\mathrm{F}}}, \\
\frac{\Gamma_2}{\Gamma} &= \frac{3}{p_{\mathrm{F}}^2}.
\end{align}\end{subequations}   Finally, \begin{align}
\mathsf{W}_{\mathrm{imp}} &\approx \Gamma \left(1+\frac{5\mpi^2 T^2}{8T_{\mathrm{F}}^2}\right)|0\rangle\langle 0| - \frac{\sqrt{3}\mpi \Gamma T}{2T_{\mathrm{F}}} (|1\rangle\langle0|+|0\rangle\langle 1|) +  \Gamma \left(1+\frac{137\mpi^2 T^2}{40T_{\mathrm{F}}^2}\right)|1\rangle\langle 1| + \frac{2\mpi^2 \Gamma T^2}{\sqrt{5}T_{\mathrm{F}}^2}(|2\rangle\langle 0|+|0\rangle\langle 2|) \notag \\
& \;\;\;\; - \sqrt{\frac{12}{5}}\frac{\mpi \Gamma T}{T_{\mathrm{F}}}(|2\rangle\langle1|+|1\rangle\langle 2|) + \Gamma  \left(1+\frac{2523\mpi^2 T^2}{280T_{\mathrm{F}}^2}\right)|2\rangle\langle 2| 
\end{align} 
Employing (\ref{eq:thermoelcondboltz}), we find (\ref{eq:34eq}).

Finally, let us justify that the relaxation time approximation for electron-electron collisions does not change the form of (\ref{eq:34eq}).  Using (\ref{eq:QxC}), the electron-electron contribution to $T\bar\kappa$ is \begin{equation}
T\bar\kappa \approx \frac{\mpi^2 T^2 v_{\mathrm{F}}^2}{3} \langle 1 | \widetilde{\mathsf{W}}^{-1}|1\rangle = \frac{\mpi^2 T^2 v_{\mathrm{F}}^2}{3\Gamma}\left[1 - \frac{1}{\Gamma} \langle 1| (\widetilde{\mathsf{W}}-\Gamma)|1\rangle + \cdots\right] = \frac{\mpi^2 T^2 v_{\mathrm{F}}^2}{3\Gamma}\left[1 - \frac{\langle 1|\mathsf{W}|1\rangle}{\Gamma}  + \cdots\right].
\end{equation}
As $\langle 1|\mathsf{W}|1\rangle$ is $\mathrm{O}(T^2)$, we conclude that so long as $\alpha^2 T^2/T_{\mathrm{F}} \equiv \langle 1|\mathsf{W}|1\rangle$, (\ref{eq:34eq}) is unchanged for arbitrary collision integrals.

\section{Thermal Conductivity in the Hydrodynamic Limit}
\label{app:kappa}
In this appendix we explicitly compute the hydrodynamic  dissipative coefficient $\kappa_{\textsc{q}}$.  In a Galilean-invariant fluid, the heat current is given by \begin{equation}
\mathbf{Q} = Ts\mathbf{v}_{\mathrm{hydro}}  - \kappa_{\textsc{q}}  \nabla T.  \label{eq:QhydroC}
\end{equation}
To see how this arises within kinetic theory, consider the following  component of the time-independent Boltzmann in  a translation-invariant fluid:  \begin{equation}
 v_j \partial_j |\mathsf{Q}_i\rangle \approx - \gamma \left(|\mathsf{Q}_i\rangle - \frac{\langle \mathsf{Q}_i|\mathsf{P}_j\rangle}{\langle  \mathsf{P} |\mathsf{P}\rangle}|\mathsf{P}_j\rangle\right)
\end{equation}   
%
In a  hydrodynamic limit,  $v_j \partial_j  |\mathsf{Q}_i\rangle $ is dominated by any vector associated with a  conserved quantity -- since these gradients are long lived modes.   Assume for simplicity that charge, energy and momentum are the only conserved quantities:  the  first two are given by \begin{subequations}\begin{align}
|\mathsf{n}\rangle &= \int  \mathrm{d}^d\mathbf{p} \; |\mathbf{p}\rangle, \\ 
|\mathsf{e}\rangle &= \int  \mathrm{d}^d\mathbf{p} \; \epsilon(\mathbf{p})|\mathbf{p}\rangle. 
\end{align}\end{subequations}
Note the identities \begin{subequations}\begin{align}
\langle \mathsf{n}|\mathsf{n} \rangle &= \int  \frac{\mathrm{d}^d\mathbf{p}}{(2\mpi\hbar)^d} \left(-\frac{\partial f_{\mathrm{eq}}}{\partial \epsilon}\right) = \frac{\partial  n}{\partial \mu} \equiv \chi, \\
\langle \mathsf{e}|\mathsf{e} \rangle &=  \int  \frac{\mathrm{d}^d\mathbf{p}}{(2\mpi\hbar)^d} \left(-\frac{\partial f_{\mathrm{eq}}}{\partial \epsilon}\right) \epsilon^2 \approx \frac{\mpi^2 T^2 \nu }{3}, \\
\langle \mathsf{n}|\mathsf{e}\rangle  &\sim T^2
\end{align}\end{subequations}
with $\chi$  the charge compressibility; note $\chi \sim T^0$.  Since 
\begin{subequations}\begin{align}
\langle \mathsf{n}| v_j |\mathsf{Q}_i\rangle &= \frac{\mdelta_{ij}}{d} \int \frac{\mathrm{d}^d\mathbf{p}}{(2\mpi\hbar)^d} \left(-\frac{\partial f_{\mathrm{eq}}}{\partial \epsilon}\right) v^2 \epsilon  \sim T^2,  \\
\langle \mathsf{e}| v_j |\mathsf{Q}_i\rangle &= \frac{\mdelta_{ij}}{d} \int \frac{\mathrm{d}^d\mathbf{p}}{(2\mpi\hbar)^d} \left(-\frac{\partial f_{\mathrm{eq}}}{\partial \epsilon}\right) v^2 \epsilon^2 \approx \langle \mathsf{e}|\mathsf{e}\rangle \frac{v_{\mathrm{F}}^2}{d}\mdelta_{ij}
\end{align}\end{subequations}
using  (\ref{eq:QPinnerproduct}), we conclude that \begin{equation}
|\mathsf{Q}_i\rangle - Ts  \frac{|\mathsf{P}_i\rangle}{\langle \mathsf{P}|\mathsf{P}\rangle} = - \frac{v_{\mathrm{F}}^2}{d\gamma} \partial_i |\mathsf{e}\rangle  + \mathrm{O}\left(T^2 \partial_i | \mathsf{n}\rangle\right)  \label{eq:QiC}
\end{equation}

In order to relate fluctuations in temperature $\mdelta T$ to $|\mathsf{e}\rangle$, observe that \begin{equation}
f_{\mathrm{eq}}(T+\mdelta T) - f_{\mathrm{eq}}(T) = \left(\frac{\epsilon}{T}\mdelta T\right)  \left(-\frac{\partial f_{\mathrm{eq}}}{\partial \epsilon}\right),
\end{equation}
which implies that \begin{equation}
\langle \mathsf{e} |\Phi\rangle = \mdelta T \times \frac{\langle \mathsf{e}|\mathsf{e}\rangle}{T} = \mdelta T \times \frac{\mpi^2 T\nu}{3}.
\end{equation}
Taking the inner product of (\ref{eq:QiC}) with a generic distribution function $|\Phi\rangle$,  and noting that the approximate low-temperature identity $\mathbf{J} = n \mathbf{v}_{\mathrm{hydro}}$ implies that $\langle \Phi|\mathsf{P}_i\rangle = \langle \mathsf{P}|\mathsf{P}\rangle v_{\mathrm{hydro},i}$, we conclude that (\ref{eq:QhydroC}) holds with \begin{equation}
\kappa_{\textsc{q}}  = \frac{\mpi^2 \nu Tv_{\mathrm{F}}^2}{3d\gamma}.
\end{equation}
This expression precisely agrees with the open-circuit thermal conductivity (\ref{eq:kappa}) in the hydrodynamic limit $\gamma \gg \Gamma$.

\end{appendix}

\bibliographystyle{unsrt}
\addcontentsline{toc}{section}{References}
\bibliography{wfbib}

\end{document}